\documentclass[review]{elsarticle}

 \usepackage{graphicx}
 \usepackage{caption}
 \usepackage{subcaption}
 \usepackage{epstopdf}
 \usepackage[percent]{overpic}
 \usepackage{stackengine}
 \usepackage{makecell}
 \usepackage{multirow}

 \usepackage{silence}
\usepackage{amssymb}
\usepackage[fleqn]{amsmath}
 \usepackage{nicematrix}
\usepackage{commath}

 \usepackage{lineno}

\biboptions{sort&compress}

\usepackage{dashrule}
\usepackage[nameinlink]{cleveref}
    \crefname{equation}{Eq.}{Eqs.}
    \crefname{figure}{Fig.}{Figs.}
    \crefname{table}{Table}{Tables}

\usepackage{url}
\usepackage{comment}
\usepackage{etoolbox}
\usepackage{booktabs}
\usepackage{todonotes}
  \presetkeys{todonotes}{noline,inline}{} 
  
\usepackage{psfrag}
\usepackage{tikz}
\usetikzlibrary{shapes}
\definecolor{colKLW}{HTML}{0072B2}
\definecolor{colEBRSM}{HTML}{CC79A7}
\usepackage{adjustbox}
\usepackage{siunitx}
\usepackage{eurosym}

\usepackage{textcomp}
\usepackage{float}
  \newfloat{nomenbox}{t}{nmcl}
\floatstyle{boxed} 
  \restylefloat{nomenbox}
\usepackage{color}
\usepackage{xcolor}
\usepackage{tabularx,ragged2e}
\usepackage{array,tabulary}
\usepackage{multicol}
\usepackage{placeins}
\usepackage{flafter}
\usepackage{wasysym}
\usepackage{xfrac}
\usepackage{mathrsfs}
\usepackage{tablefootnote}

\usepackage{pgfplots, pgfplotstable}
\pgfplotsset{compat=1.9}
\usepackage[normalem]{ulem}

\usepackage{lscape} % for landscape environment
\usepackage{longtable}
\usepackage{threeparttable}

\usepackage{appendix}
\usepackage{lmodern}
\usepackage{svg}

\usepackage{comment}

\newcommand{\Pra}{\mathrm{Pr}}

\usepackage{nomencl} 	% nomenclature definitions
\makenomenclature	% This has to be in the preamble
\RequirePackage{ifthen} 
\renewcommand{\nomgroup}[1]{%
\ifthenelse{\equal{#1}{A}}{\item[\emph{Roman letters}]}{%{}}}
\ifthenelse{\equal{#1}{G}}{\item[\emph{Greek letters}]}{%{}}}
\ifthenelse{\equal{#1}{N}}{\item[\emph{Non-dimensional numbers}]}{%
\ifthenelse{\equal{#1}{S}}{\item[\emph{Subscripts}]}{%
\ifthenelse{\equal{#1}{O}}{\item[\emph{Operators}]}{%
\ifthenelse{\equal{#1}{P}}{\item[\emph{Acronyms}]}{%%}}}}
}}}}}}}

\usetikzlibrary{arrows.meta}
\newcommand{\keps}{$k$-$\epsilon$}

\newcommand{\kwsst}{$k$-$\omega$ SST}
\newcommand{\kThetaEpsTheta}{$k_\theta$-$\epsilon_\theta$}
\newcommand{\kThetaOmegaTheta}{$k_\theta$-$\Omega_\theta$}

\newcommand{\daVia}{$k$-$\Omega$-$k_\theta$-$\Omega_\theta$}

\newcommand{\dissRate}{\varepsilon = \nu\overline{\dx{u^\prime_i}{x_j}\dx{u^\prime_i}{x_j}}}

\newcommand{\ofversion}{\mbox{OpenFOAM v2212}}

\newlength{\figureWidth}

\renewcommand\sout{\bgroup\markoverwith
{\textcolor{red}{\rule[0.5ex]{2pt}{0.4pt}}}\ULon}

\newcommand{\dx}[2]{\frac{\partial{#1}}{\partial{#2}}}

\newcommand{\simFalseSmall}{%
    \begin{tikzpicture}baseline={(current bounding box.north)}
        \node[draw,circle,black,minimum size=1pt,scale=0.3] at (0,0) {};
    \end{tikzpicture}%
}
\graphicspath{ {./} }

\journal{International Journal of Heat and Fluid Flow}
\newcolumntype{R}{@{\extracolsep{1cm}}r@{\extracolsep{0pt}}}%
\begin{document}
\sloppy
\setlength{\figureWidth}{\columnwidth}

\renewcommand{\epsilon}{\varepsilon}

\begin{frontmatter}


\title{Critical assessment  of RANS Models for Turbulent Heat Transfer in Low-Prandtl-Number Forced Convection}


\author[PM]{L.~Marocco}
\author[KIT]{J.~Schmitt}
\author[KIT]{J.~Neuhauser}
\author[KIT]{B.~Frohnapfel}
\author[KIT]{D.~Gatti}
% \author[PM]{L.~Marocco}
\ead{davide.gatti@kit.edu}
% \ead{luca.marocco@polimi.it}
\address[KIT]{Karlsruhe Institute of Technology, Institute of Fluid Mechanics, Kaiserstr. 10, Karlsruhe, Germany}
\address[PM]{Politecnico di Milano, Department of Energy, via Lambruschini 4, Milano, Italy}
\cortext[cor1]{Corresponding author}

% \corref{cor2}}

\begin{abstract}
% ===== begin abstract.tex =====
Reliable modeling of turbulent
heat transfer in liquid metals via Reynolds--Averaged Navier--Stokes (RANS) remains challenging because the very low
Prandtl number weakens the similarity between momentum and thermal
transport. In this work, several thermal turbulence closures for forced
convection in liquid metals are assessed using the finite-volume code
OpenFOAM v2212. The investigated model combinations include 
the $k$--$\omega$ SST model with the Kays
correlation for the turbulent Prandtl number, the four-equation
$k$--$\epsilon$--$k_\theta$--$\epsilon_\theta$ model, the logarithmic
$k$--$\Omega$--$k_\theta$--$\Omega_\theta$ model, and two algebraic
heat-flux formulations coupled with either a low-Reynolds $k$--$\epsilon$
model or an elliptic blending Reynolds-stress model (EBRSM). The models are
evaluated in turbulent channel flow, pipe flow, and heated
backward-facing step flow against direct numerical simulation data and
reference results reported in the original model publications, with
particular attention to reproducibility, numerical robustness, and
predictive accuracy.

The results show that only a limited subset of models can be considered
reliable for low-Prandtl-number flows. The  $k$--$\omega$ SST 
combined with the Kays correlation provides the most robust overall
performance and accurate temperature predictions in all cases. The
$k$--$\epsilon$--$k_\theta$--$\epsilon_\theta$ model also shows good
reproducibility and satisfactory agreement with reference data,
remaining the most consistent multi-equation alternative. In contrast,
the logarithmic four-equation model exhibits reduced numerical
robustness, while the algebraic heat-flux model coupled with the
$k$--$\epsilon$ momentum closure fails to reproduce published thermal
results despite correct prediction of the momentum field. The
EBRSM-based algebraic heat-flux formulation reproduces selected
reference results but suffers from significant robustness limitations.
The study establishes a unified formulation of the examined closures by
reconciling discrepancies and correcting inconsistencies in their
published formulations, and systematically verifies their
reproducibility and robustness in low-Prandtl-number forced convection.
% ===== end abstract.tex =====
\end{abstract}

\begin{keyword}
RANS \sep Liquid metals \sep Low-Prandtl-number flows \sep
Turbulent heat transfer \sep Reproducibility \sep
Numerical robustness
\end{keyword}

\end{frontmatter}
%% Line numbering starts here

% Nomenclature
% \input{nomenclature2}
% \clearpage
% \input{nomenclature}
% \renewcommand{\nomname}{Nomenclature}
% \setlength{\nomlabelwidth}{3cm}
% \setlength{\nomitemsep}{0pt}
% \printnomenclature[3cm]

\begin{nomenbox*}[htb!]
  \begin{multicols}{2}
  \small{

\begin{thenomenclature} 
\nomgroup{A}
  \item [{$c_f$}]\begingroup skin-friction coefficient\nomeqref {0}\nompageref{3}
  \item [{$c_p$}]\begingroup specific heat capacity at constant pressure\nomeqref {0}\nompageref{3}
  \item [{$d_w$}]\begingroup distance to the nearest wall\nomeqref {0}\nompageref{3}
  \item [{$D$}]\begingroup pipe diameter\nomeqref {0}\nompageref{3}
  \item [{$ER$}]\begingroup expansion ratio of the backward-facing step\nomeqref {0}\nompageref{3}
  \item [{$k$}]\begingroup turbulent kinetic energy\nomeqref {0}\nompageref{3}
  \item [{$l_{\mathrm{step}}$}]\begingroup backward-facing step height\nomeqref {0}\nompageref{3}
  \item [{$k_\theta$}]\begingroup half temperature variance\nomeqref {0}\nompageref{3}
  \item [{$Nu_x$}]\begingroup local Nusselt number\nomeqref {0}\nompageref{3}
  \item [{$p$}]\begingroup pressure\nomeqref {0}\nompageref{3}
  \item [{$P_k$}]\begingroup production of turbulent kinetic energy\nomeqref {0}\nompageref{3}
  \item [{$P_\theta$}]\begingroup thermal production term\nomeqref {0}\nompageref{3}
  \item [{$Pr$}]\begingroup Prandtl number\nomeqref {0}\nompageref{3}
  \item [{$Pr_t$}]\begingroup turbulent Prandtl number\nomeqref {0}\nompageref{3}
  \item [{$q_w$}]\begingroup wall heat flux\nomeqref {0}\nompageref{3}
  \item [{$R$}]\begingroup time-scale ratio\nomeqref {0}\nompageref{3}
  \item [{$Re$}]\begingroup Reynolds number\nomeqref {0}\nompageref{3}
  \item [{$Re_b$}]\begingroup bulk Reynolds number\nomeqref {0}\nompageref{3}
  \item [{$Re_\tau$}]\begingroup friction Reynolds number\nomeqref {0}\nompageref{3}
  \item [{$T$}]\begingroup temperature\nomeqref {0}\nompageref{3}
  \item [{$T_{\mathrm{ref}}$}]\begingroup reference temperature\nomeqref {0}\nompageref{3}
  \item [{$T_\tau$}]\begingroup friction temperature\nomeqref {0}\nompageref{3}
  \item [{$u_i$}]\begingroup velocity components in Cartesian coordinates\nomeqref {0}\nompageref{3}
  \item [{$u_b$}]\begingroup bulk velocity\nomeqref {0}\nompageref{3}
  \item [{$u_\tau$}]\begingroup friction velocity\nomeqref {0}\nompageref{3}
  \item [{$x$, $y$, $z$}]\begingroup Cartesian coordinates\nomeqref {0}\nompageref{3}
\nomgroup{G}
  \item [{$\alpha$}]\begingroup thermal diffusivity\nomeqref {0}\nompageref{3}
  \item [{$\alpha_t$}]\begingroup turbulent thermal diffusivity\nomeqref {0}\nompageref{3}
  \item [{$\delta$}]\begingroup channel half-height\nomeqref {0}\nompageref{3}
  \item [{$\delta_{ij}$}]\begingroup Kronecker delta\nomeqref {0}\nompageref{3}
  \item [{$\epsilon$}]\begingroup pseudo dissipation rate of $k$\nomeqref {0}\nompageref{3}
  \item [{$\epsilon_\theta$}]\begingroup dissipation rate of $k_\theta$\nomeqref {0}\nompageref{3}
  \item [{$\theta$}]\begingroup temperature variable\nomeqref {0}\nompageref{3}
  \item [{$\lambda$}]\begingroup thermal conductivity\nomeqref {0}\nompageref{3}
  \item [{$\nu$}]\begingroup kinematic viscosity\nomeqref {0}\nompageref{3}
  \item [{$\nu_t$}]\begingroup turbulent viscosity\nomeqref {0}\nompageref{3}
  \item [{$\rho$}]\begingroup density\nomeqref {0}\nompageref{3}
  \item [{$\tau_w$}]\begingroup wall shear stress\nomeqref {0}\nompageref{3}
  \item [{$\Omega$}]\begingroup logarithmic form of $\omega$\nomeqref {0}\nompageref{3}
  \item [{$\Omega_\theta$}]\begingroup logarithmic form of $\omega_\theta$\nomeqref {0}\nompageref{3}
  \item [{$\omega$}]\begingroup specific dissipation rate of $k$\nomeqref {0}\nompageref{3}
  \item [{$\omega_\theta$}]\begingroup specific dissipation rate of $k_\theta$\nomeqref {0}\nompageref{3}
\nomgroup{O}
  \item [{$(\cdot)'$}]\begingroup fluctuating quantity\nomeqref {0}\nompageref{3}
  \item [{$(\cdot)^+$}]\begingroup quantity scaled in wall units\nomeqref {0}\nompageref{3}
\nomgroup{P}
  \item [{AHFM}]\begingroup Algebraic Heat Flux Model\nomeqref {0}\nompageref{3}
  \item [{BFS}]\begingroup Backward-Facing Step\nomeqref {0}\nompageref{3}
  \item [{EBRSM}]\begingroup Elliptic Blending Reynolds-Stress Model\nomeqref {0}\nompageref{3}
  \item [{RANS}]\begingroup Reynolds-averaged Navier--Stokes\nomeqref {0}\nompageref{3}
  \item [{SGDH}]\begingroup Simple Gradient Diffusion Hypothesis\nomeqref {0}\nompageref{3}

\end{thenomenclature}
    }
  \end{multicols}
\end{nomenbox*}
\clearpage

\FloatBarrier

\section{Introduction} 
\label{sec:intro}
% ===== begin introduction.tex =====
The increasing demand for efficient thermal management in advanced
energy systems has renewed interest in liquid metals as heat transfer
fluids. The ratio between momentum and thermal diffusivity, the so-called Prandtl number ($\Pra$), is in the range $10^{-3}$--$10^{-2}$ for liquid metals and thus very small, thus decoupling momentum and thermal transport
\cite{grotzbach2013,Straub2019,Marocco2016}. As a consequence, standard
turbulence modeling assumptions, such as the similarity between
velocity and temperature fields and the use of a constant turbulent
Prandtl number, become questionable \cite{mathur2023}.

% Most turbulence models currently employed in Reynolds-averaged
% Navier--Stokes (RANS) simulations were originally developed and
% calibrated for fluids with Prandtl numbers of order unity
% ($Pr \approx 1$). Their application to liquid-metal flows therefore
% requires extensions beyond their original range of validity, especially
% for the modeling of turbulent heat transfer.

In the Reynolds-Averaged Navier---Stokes (RANS) framework, turbulent heat transfer is modeled through two
distinct components: a momentum turbulence model, which provides the
Reynolds stresses, either through the Boussinesq hypothesis in terms of
a turbulent viscosity or via Reynolds-stress transport equations, and
a thermal turbulence model, which determines the turbulent heat flux
through an appropriate closure.
% These two components play fundamentally
% different roles and must be clearly distinguished.

Momentum turbulence models are
independent of the fluid $\Pra$ number. The most commonly adopted formulations for momentum turbulence modeling
are based on high-Reynolds-number $k$--$\epsilon$ models with wall
functions, as well as low-Reynolds-number models of the $k$--$\omega$
family, in particular the widely used $k$--$\omega$ SST model
\cite{Menter2003}. More advanced closures include low-Reynolds-number
formulations such as the $k$--$\epsilon$ models  of \citet{AbeKondoh1994,Lien1996}, and Reynolds-stress models such as
the elliptic blending Reynolds-stress model (EBRSM) of 
\citet{Manceau2014}. These models provide the turbulent viscosity and
constitute the basis for the momentum turbulence component of all model
combinations investigated in the present study.

In contrast, thermal turbulence models are directly affected by the
Prandtl number, as they govern the relationship between momentum and
thermal transport. Most available thermal closures have been originally
developed for fluids with $Pr \approx 1$, where the analogy between velocity and temperature
fields is reasonably valid. This includes temperature-variance-based
models derived from the work of Abe et al. \cite{abe1993} and algebraic
heat-flux formulations based on Kenjeres et al.
\cite{Kenjeres2005}.

Extensions of these models to low-Prandtl-number flows have been
proposed through modified formulations, additional terms, or empirical
corrections. However, these extensions are not uniquely defined and
often differ across publications, leading to ambiguities in the
mathematical formulation and difficulties in reproducing published
results. As a consequence, discrepancies between different studies are
frequently observed, even when nominally the same model is employed. 

Within this context, the present work focuses on a set of thermal
turbulence closures that have been widely adopted and further developed
for low-Prandtl-number flows. These include multi-equation models based
on temperature variance transport, such as the four-equation
$k$--$\epsilon$--$k_\theta$--$\epsilon_\theta$ model proposed by
\citet{Manservisi2014} and the logarithmic
$k$--$\Omega$--$k_\theta$--$\Omega_\theta$ model introduced by
\citet{Manservisi2016}, both derived from the original formulations of
\citet{abe1993}. In addition, algebraic heat-flux models
(AHFM), such as those proposed by \citet{Shams2014}, are considered.
These models are based on the formulation of 
\citet{Kenjeres2005} and have been extended to liquid-metal flows and
coupled with different momentum turbulence models, including
low-Reynolds-number closures derived from  \citet{Lien1996} and
Reynolds-stress models such as the EBRSM of \citet{Manceau2014}.

A simpler alternative consists in modeling the turbulent heat flux
through a turbulent Prandtl number, as in the correlation proposed by
Kays \cite{Kays1994}, here coupled with a $k$--$\omega$ SST
momentum turbulence model \cite{Menter2003}. This approach provides a
local relation between turbulent viscosity and thermal diffusivity.
While computationally efficient, it remains empirical and its validity
in low-Prandtl-number flows is not guaranteed a priori.

The present study addresses these issues through a systematic
assessment of RANS turbulence models for forced convection in liquid
metals with three main objectives: (i) to reconstruct and verify the
mathematical formulation of selected momentum and thermal turbulence
models by resolving discrepancies, ambiguities, and inconsistencies
found in their original publications, (ii) to assess the
reproducibility of reference results reported by the model developers,
and (iii) to evaluate model accuracy against high-fidelity data,
including direct numerical simulation (DNS) and large-eddy simulation
(LES). To the authors’ knowledge, these aspects have not been addressed
simultaneously in a unified framework for low-Prandtl-number flows. The assessment deliberately targets
the low-Prandtl-number regime ($Pr\le 0.025$) for which the considered
closures and their coefficients were calibrated; their parent formulations
for $Pr\approx 1$ fluids have been extensively validated in the original
publications \cite{abe1993,Kenjeres2005,Kays1994}.

The model families considered in this work have been selected based on their
widespread use and continued development in the literature \cite{roelofs2020}. However, as detailed
in \Cref{sec:turbulenceModels}, the specific published forms of these models contain discrepancies across
publications. Reconstructing a single consistent formulation, and documenting the
choices made, is one of the objectives of the present work.

% Particular care has been devoted to the
% verification of governing equations, model coefficients, and boundary
% conditions, as inconsistencies and omissions in the literature can
% significantly affect model behavior and lead to misleading conclusions.

The assessment is carried out on a set of canonical and complex test
cases, including turbulent channel flow, pipe flow, and heated
backward-facing step flow. These configurations are commonly adopted in the original model development studies. Moreover, they are standard benchmarks
in the literature and are characterized by the availability of
high-fidelity reference data from DNS
and LES, allowing for a systematic evaluation
of predictive accuracy, numerical robustness, and reproducibility
across different flow regimes and model combinations.

All models have been implemented into  (and tested with)  a single, finite-volume code \mbox{OpenFOAM
v2212} \cite{OpenFOAMv2212}, adopting a modular framework in which
momentum and thermal turbulence closures are treated independently and
can be combined systematically. 
Furthermore, the implemented models, the corresponding simulation
cases, and a complete database of the obtained results are provided as
accompanying material. This guarantees full reproducibility of the
present study, enables consistent and unbiased comparisons across
different modeling approaches, and establishes a reference dataset for
future development and validation of turbulence models for
low-Prandtl-number flows. 

The outline of the paper is as follows. \Cref{sec:turbulenceModels} introduces the considered turbulence models in detail, including their formulation. In \cref{sec:numericalSetup}, the numerical setup of the test cases is described, which are used for model assessment in \cref{sec:resultsAndDiscussion} together with a discussion of the reproducibility and numerical stability for each of the models.
% ===== end introduction.tex =====

\section{Considered turbulence models and their formulation}
\label{sec:turbulenceModels}
% ===== begin turbulenceModels.tex =====
This section presents the turbulence models considered in the present
study together with their mathematical formulation. For clarity, the
models are organized according to their role in the closure of the
RANS equations, distinguishing between momentum turbulence models and
thermal turbulence models.

The momentum turbulence models are described first, as they provide the
underlying closure for the Reynolds stresses and supply the quantities
required by the thermal turbulence models. The thermal turbulence
models are then introduced separately, grouped according to the type of
closure adopted for the turbulent heat flux, namely turbulent-Prandtl-number,
multi-equation, and algebraic heat-flux approaches.

Although thermal turbulence models depend on the underlying momentum
closure, this separation provides a clearer presentation of the model
hierarchy and of the specific role of each component. Their actual
coupling in the simulations is defined through the model combinations
summarized in \Cref{tab:turb_models} and discussed in
\Cref{subsec:model_selection}.

\subsection{Model selection}
\label{subsec:model_selection}

The turbulence model combinations considered in this study are
summarized in \Cref{tab:turb_models}. They are representative of the
most widely adopted approaches for modeling turbulent heat transfer in
low-$\Pra$-number flows.

Each combination consists of a momentum turbulence model and a thermal
turbulence closure. For the multi-equation thermal closures, it should
be noted that the thermal model is calibrated together with a specific
momentum turbulence model. Therefore, combining such closures with a
different momentum model generally leads to poor performance. 

An alternative assessment strategy consists in combining different thermal
closures with a single, reliable momentum model, which would isolate the intrinsic
accuracy of the thermal closures. This approach is deliberately not pursued here: the
multi-equation thermal models are calibrated jointly with their momentum closure, and
departing from the published pairings would compromise the reproducibility assessment,
which is the primary objective of this study. The modular framework provided as
supplementary material nevertheless enables such cross-combination studies, which
represent a natural extension of the present work.

The abbreviations introduced in \Cref{tab:turb_models} are used
consistently throughout the paper.

\begin{table}
\centering
\caption{Overview of selected turbulence model combinations. The first model in each block corresponds to the momentum turbulence model, and the second represents the thermal turbulence model.}
\begin{tabularx}{\textwidth}{X X X}
\toprule
\textbf{Momentum Model} & \textbf{Thermal Model} & \textbf{Abbreviation} \\
\midrule 
\kwsst & Kays Correlation & KWSST-KAYS \\
\midrule 
\keps{} (AKN) & \kThetaEpsTheta (MM) & AKN-MM \\
\midrule
$k$-$\Omega$  (KLW) & \kThetaOmegaTheta (DAVIA) & KLW-DAVIA \\
\midrule
\keps{} (ShamsKE) & AHFM-NRG & ShamsKE-AHFM \\
\midrule
EBRSM & AHFM-NRG & EBRSM-AHFM \\
\bottomrule
\end{tabularx}
\label{tab:turb_models}
\end{table}

\subsection{Momentum turbulence models}
\label{subsec:momentum_turb_models}
The momentum turbulence models considered in this study are those
forming the basis of the thermal turbulence closures investigated in
the following.

In particular, the low-Reynolds-number $k$--$\epsilon$ model of
\citet{AbeKondoh1994} (AKN) and its logarithmic reformulation as a
$k$--$\Omega$ model proposed by \citet{Manservisi2016} (KLW) provide
the underlying momentum closures for the multi-equation thermal
turbulence models MM and DAVIA, respectively.

For the algebraic heat-flux approach, the AHFM closure is coupled with
either the ShamsKE variant proposed by \citet{Shams2014}, based on the
low-Reynolds-number $k$--$\epsilon$ model of \citet{Lien1996}, or with
the EBRSM of \citet{Manceau2014}.

The $k$--$\omega$ SST model \cite{Menter2003} and the EBRSM
\cite{Manceau2014,Manceau2015} are employed as
implemented in \mbox{OpenFOAM v2212} \cite{OpenFOAMv2212}. For these models,
reference is made to \citet{Manceau2014,Manceau2015} and to the OpenFOAM
documentation. 

In the present study, the EBRSM is used as implemented in
\ofversion{}, following the formulation of \citet{Manceau2015}. This
differs from the version considered by \citet{Shams2019Number3}, which
is based on \citet{Manceau2014}. The use of different implementations
of the EBRSM may therefore represent one possible source of the
discrepancies observed in the present assessment.

The AKN and KLW models, and the ShamsKE formulation are described in detail in
the following subsections.

\subsubsection{$k$--$\epsilon$ AKN}
\label{subsubsec:akn}

The low-Reynolds-number $k$--$\epsilon$ model proposed by
\citet{AbeKondoh1994} is used as the reference momentum turbulence
model for the multi-equation thermal closures ($k_\theta$-$\epsilon_\theta$ and $k_\theta$-$\Omega_\theta$)  considered in this study.
The model is formulated in terms of transport equations for the
turbulent kinetic energy $k$ and its pseudo-dissipation rate
$\epsilon$:
\begin{align}
\label{eqn:akn_k}
\frac{\partial k}{\partial t} + u_i \frac{\partial k}{\partial x_i} &=
\frac{\partial}{\partial x_i}
\left[
\left( \nu + \frac{\nu_t}{\sigma_k} \right)
\frac{\partial k}{\partial x_i}
\right] + P_k - \epsilon  \\
\label{eqn:akn_eps}
\frac{\partial \epsilon}{\partial t} + u_i \frac{\partial \epsilon}{\partial x_i} &=
\frac{\partial}{\partial x_i}
\left[
\left( \nu + \frac{\nu_t}{\sigma_\epsilon} \right)
\frac{\partial \epsilon}{\partial x_i}
\right]
+ C_{\epsilon1} \frac{\epsilon}{k} P_k
- C_{\epsilon2} f_\epsilon \frac{\epsilon^2}{k} 
\end{align}
The production of turbulent kinetic energy is defined as
\begin{equation}
\label{eqn:prodK}
P_k = -\overline{u_i' u_j'} \frac{\partial u_i}{\partial x_j} \, ,
\end{equation}
while the Reynolds stresses $\overline{u_i' u_j'}$ are modeled using the Boussinesq hypothesis
\begin{equation}
\label{eqn:re_stress}
\overline{u_i' u_j'} =
- \nu_t
\left(
\frac{\partial u_i}{\partial x_j} +
\frac{\partial u_j}{\partial x_i}
\right)
+ \frac{2}{3} k \delta_{ij} 
\end{equation}

The quantity $\varepsilon$ represents the pseudo-dissipation rate \cite{Pope_2000}, defined as
\begin{equation}
\label{eqn:epsAKN}
    \dissRate \,
\end{equation}
Within the model, $\varepsilon$ is obtained from its transport equation (\Cref{eqn:akn_eps}). \Cref{eqn:epsAKN}
only specifies the quantity being modeled, which is relevant for the formulation of
the wall boundary condition and for the comparison with DNS data.

% The pseudo dissipation rate has the form \cite{Pope_2000}
% \begin{equation}
% \label{eqn:epsAKN}
%     \dissRate \,
% \end{equation}
The turbulent viscosity is defined as
\begin{equation}
\label{eqn:nu_t}
\nu_t = C_\mu f_\mu \frac{k^2}{\epsilon} 
\end{equation}
The model coefficients in \Cref{eqn:akn_k,eqn:akn_eps,eqn:nu_t} are
\begin{equation}
\label{eqn:akn_coeff}
\sigma_k = 1.4, \quad
\sigma_\epsilon = 1.4, \quad
C_{\epsilon1} = 1.5, \quad
C_{\epsilon2} = 1.9, \quad
C_\mu = 0.09 
\end{equation}
while the damping functions
$f_\mu$ and $f_\epsilon$, with which the near-wall effects are accounted for, are defined as
\begin{align}
\label{eqn:fmu}
f_\mu &=
\left(1 - e^{-R_\delta / 14}\right)^2
\left(
1 + \frac{5}{R_t^{3/4}} e^{-(R_t/200)^2}
\right) \\
\label{eqn:feps}
f_\epsilon &=
\left(1 - e^{-R_\delta / 3.1}\right)^2
\left(1 - 0.3 e^{-(R_t/6.5)^2}\right)
\end{align}
Here, $R_t$ is the turbulent Reynolds number and $R_\delta$ is the ratio of the distance to the nearest wall to the Kolmogorov length scale. They are defined as 
\begin{equation}
\label{eqn:Rt_Rdelta}
R_t = \frac{k^2}{\nu \epsilon} \qquad \text{and} \qquad 
R_\delta = \frac{d_w}{\left( \nu^3 / \epsilon \right)^{1/4}}
\end{equation}
with $d_w$ denoting the distance to the nearest wall.

In the original formulation of \citet{AbeKondoh1994}, near-wall
scaling is expressed in terms of wall units ($y^+$). In the present
work, a fully local formulation based on $R_t$ and $R_\delta$ is
adopted, following \citet{Manservisi2014}. This avoids the explicit
evaluation of wall units and ensures consistency when applying the
model to complex geometries and unstructured meshes.

\subsubsection{$k$--$\Omega$ model (KLW)}
\label{subsubsec:keOmega}

The $k$--$\Omega$ model was introduced by \citet{Manservisi2016} as the
momentum turbulence closure within the \daVia{} model combination. The
model is based on the formulation of \citet{AbeKondoh1994}, with the
dissipation rate $\epsilon$ replaced by the logarithmic variable
$\Omega$, defined as:
\begin{equation}
\label{eqn:Omega}
\Omega = \ln(\omega) \qquad \text{where} \quad\omega = \frac{\epsilon}{C_\mu k} \quad \text{and} \quad  C_\mu = 0.09 
\end{equation}
The use of a logarithmic transformation of the specific dissipation
rate $\omega$ is intended to improve numerical robustness compared to the AKN
model, as discussed by \citet{Manservisi2016}.
The transport equation for $k$ is identical to
\Cref{eqn:akn_k}, with the substitution $\epsilon = C_\mu k e^\Omega$, which must also be applied in \Cref{eqn:nu_t,eqn:Rt_Rdelta}.

The KLW model differs from the AKN model by solving the following
transport equation for $\Omega$ instead of $\epsilon$:
\begin{equation}
\label{eqn:OmegaKLW}
\begin{split}
\frac{\partial \Omega}{\partial t} + u_i \frac{\partial \Omega}{\partial x_i} &=
\frac{\partial}{\partial x_i}
\left[
\left( \nu + \frac{\nu_t}{\sigma_{\epsilon}} \right)
\frac{\partial \Omega}{\partial x_i}
\right]
+ \frac{2}{k}
\left( \nu + \frac{\nu_t}{\sigma_{\epsilon}} \right)
\frac{\partial k}{\partial x_i}
\frac{\partial \Omega}{\partial x_i} \\
&\quad +
\left( \nu + \frac{\nu_t}{\sigma_{\epsilon}} \right)
\frac{\partial \Omega}{\partial x_i}
\frac{\partial \Omega}{\partial x_i}
+ \frac{C_{\epsilon1} - 1}{k} P_k
- C_\mu (C_{\epsilon2} f_\epsilon - 1) e^\Omega 
\end{split}
\end{equation}
The model coefficients $\sigma_k$, $\sigma_\epsilon$, $C_{\epsilon1}$,
and $C_{\epsilon2}$, as well as the damping functions $f_\mu$ and
$f_\epsilon$, are identical to those of the AKN model
(\Cref{subsubsec:akn}).
The second and third terms on the right-hand side of
\Cref{eqn:OmegaKLW} correspond to cross-diffusion contributions. By
introducing
\[
D_1 = \frac{2}{k}\left( \nu + \frac{\nu_t}{\sigma_{\epsilon}} \right),
\qquad
D_2 = \left( \nu + \frac{\nu_t}{\sigma_{\epsilon}} \right)
\]
they can be rewritten as:
\begin{align*}
D_1 \frac{\partial k}{\partial x_i} \frac{\partial \Omega}{\partial x_i}
&=
\frac{\partial}{\partial x_i}
\left(
D_1 \frac{\partial k}{\partial x_i} \Omega
\right)
-
\Omega
\frac{\partial}{\partial x_i}
\left(
D_1 \frac{\partial k}{\partial x_i}
\right) \\
D_2 \frac{\partial \Omega}{\partial x_i} \frac{\partial \Omega}{\partial x_i}
&=
\frac{\partial}{\partial x_i}
\left(
D_2 \frac{\partial \Omega}{\partial x_i} \Omega
\right)
-
\Omega
\frac{\partial}{\partial x_i}
\left(
D_2 \frac{\partial \Omega}{\partial x_i}
\right)
\end{align*}
This reformulation enables a partially implicit treatment of the cross-diffusion terms. In the present implementation, both implicit and
explicit treatments are supported. When the implicit option is
disabled, these terms are treated as explicit source contributions.

\subsubsection{Shams $k$--$\epsilon$ model (ShamsKE)}
\label{subsubsec:shamske}

The ShamsKE model is a low-Reynolds-number turbulence model used in
combination with the AHFM-NRG thermal turbulence closure
(\Cref{subsubsec:ahfm-nrg}), as introduced by \citet{Shams2014}. The
implementation adopted in this work is based on the formulation
reported by \citet{Shams2014}, which itself is derived from the
low-Reynolds-number $k$--$\epsilon$ model of \citet{Lien1996}. Since \citet{Shams2014} do not explicitly define the dissipation rate
of $k$, it is assumed that the pseudo-dissipation rate $\epsilon$
defined in \Cref{eqn:epsAKN} is employed.

The transport equations for $k$ and $\epsilon$ are:
\begin{align}
\label{eqn:kShamsKE}
\frac{\partial (\rho k)}{\partial t} + u_i \frac{\partial (\rho k)}{\partial x_i} &=
\frac{\partial}{\partial x_i}
\left[
\rho \left( \nu + \frac{\nu_t}{\sigma_k} \right)
\frac{\partial k}{\partial x_i}
\right]
+ \rho P_k - \rho \epsilon \\
\label{eqn:epsilonShamsKE}
\begin{split}
\frac{\partial (\rho \epsilon)}{\partial t} + u_i \frac{\partial (\rho \epsilon)}{\partial x_i} &=
\frac{\partial}{\partial x_i}
\left[
\rho \left( \nu + \frac{\nu_t}{\sigma_{\epsilon}} \right)
\frac{\partial \epsilon}{\partial x_i}
\right]
+ \frac{\rho C_{\epsilon1}}{\tau}(P_k + P_{\text{wall}})+ \\
&\quad - \rho C_{\epsilon2} f_\epsilon \frac{\epsilon}{\tau}
+ \rho S_Y
\end{split}
\end{align}
The production term $P_k$ is defined in \Cref{eqn:prodK}. The turbulent viscosity and auxiliary functions are:
\begin{align}
\label{eqn:nutShamsKE}
\nu_t &= C_\mu f_\mu k \tau \,, \\
\label{eqn:fmuShamsKE}
f_\mu &= 1 - \exp\left[-\left( C_{d_0}\sqrt{Re_d} + C_{d_1}Re_d + C_{d_2}Re_d^2 \right)\right] \,,\\
\label{eqn:Red}
Re_d &= \frac{\sqrt{k}\, d_w}{\nu} \,,\\
\label{eqn:fepsilonShamsKE}
f_\epsilon &= \left( 1 - C e^{-R_t^2} \right) \,,\\
\label{eqn:tauShamsKE}
\tau &= \max\left( \frac{k}{\epsilon}, \sqrt{\frac{\nu}{\epsilon}} \right)
\end{align}
where $d_w$ denotes the distance to the nearest wall.

The near-wall production term reads
\begin{equation}
\label{eqn:PwallShamsKE}
P_{\text{wall}} = D f_\epsilon
\left(P + 2\nu \frac{k}{d_w^2} \right) 
e^{-E Re_d^2} \, ,
\end{equation}
while the term $S_Y$ represents the Yap correction \cite{Yap1987}, as adopted
by \citet{Shams2018Number2}:
\begin{equation}
\label{eqn:Yap}
S_Y = C_w \frac{\epsilon}{\tau}
\max \left\{
\left( \frac{\hat{l}}{l_\epsilon} - 1 \right)
\left( \frac{\hat{l}}{l_\epsilon} \right)^2, 0
\right\},
\quad
\hat{l} = \frac{k^{3/2}}{\epsilon}, \;
l_\epsilon = C_l d_w
\end{equation}

Buoyancy effects are neglected in the present study, as only forced
convection is considered. Consequently, all buoyancy-related terms
are omitted from \Cref{eqn:kShamsKE,eqn:epsilonShamsKE}.
The model coefficients employed throughout Eqs. \eqref{eqn:kShamsKE}--\eqref{eqn:Yap} are:
\begin{align}
\label{eqn:coefficientsShamsKE}
&\sigma_k = 1, \quad \sigma_\epsilon = 1.3, \quad
C_{\epsilon1} = 1.44, \quad C_{\epsilon2} = 1.92, \quad C_\mu = 0.09, \notag \\
&C_{d_0} = 0.091, \quad C_{d_1} = 0.0042, \quad
C_{d_2} = 0.00011, \quad C = 0.3, \quad D = 1, \notag \\
&E = 0.00375, \quad C_l = 2.55, \quad C_w = 0.83
\end{align}

The implemented formulation differs from \citet{Shams2014} in two main aspects. First,  the sign in front of $C$ and the
exponential argument in \Cref{eqn:fepsilonShamsKE} are modified so that $f_\epsilon$ is consistent
with the formulation of \citet{Lien1996}. Second, the Yap correction term $S_Y$ is included following
\citet{Shams2018Number2}, with corrections to inconsistencies in the published equations. In particular, herein, the factor $1/\tau$ is retained in
\Cref{eqn:epsilonShamsKE}, the definition of $S_Y$ is expressed in
terms of $\epsilon/\tau$ rather than $\epsilon^2/k$, and the argument
of the $\max$ operator in \Cref{eqn:Yap} is formulated using 0 instead
of 1.

\subsection{Thermal turbulence models}
\label{subsec:thermal_turb_models}

Thermal turbulence models provide a closure for the turbulent heat
flux and govern the coupling between momentum and thermal transport.
In contrast to momentum turbulence models, their formulation is
directly influenced by the Prandtl number and often derived from
approaches originally developed for $Pr \approx 1$.
The models considered in this study can be grouped into three main
categories: (i) diffusivity-based approaches relying on a turbulent
Prandtl number $\Pra_t$, (ii) multi-equation models based on the transport of
temperature variance and relative dissipation rate, and 
(iii) algebraic heat-flux models providing a
direct closure for the turbulent heat flux.

The following subsections describe the specific formulations adopted
in the present work, including the Kays correlation (KAYS), the
multi-equation models MM ($k_\theta$-$\epsilon_\theta$) and 
DAVIA ($k_\theta$-$\Omega_\theta$), and the algebraic heat-flux model.

\subsubsection{Turbulent Prandtl number model (KAYS)}
\label{subsubsec:kays}

The simplest approach to model the turbulent heat flux is based on a
turbulent Prandtl number $\Pra_t$, relating the turbulent thermal diffusivity
to the turbulent viscosity through:
\begin{equation}
\label{eqn:alpha_t}
\alpha_t = \frac{\nu_t}{Pr_t} \, .
\end{equation}
In this study, the correlation proposed by \citet{Kays1994} is
employed. This semi-empirical formulation provides a local expression
for the turbulent Prandtl number as a function of the molecular
Prandtl number and the turbulent viscosity ratio:
\begin{equation}
\label{eqn:prtKays}
Pr_t = 0.85 + \frac{0.7}{Pr \, (\nu_t / \nu)} 
\end{equation}
The correlation above represents a purely local closure for
the turbulent heat flux, since it  depends explicitly on the local value of the
turbulent viscosity $\nu_t$.
The turbulent heat flux is then obtained from the diffusivity-based
closure defined as:
%in \Cref{eqn:thf}:
\begin{equation}
\label{eqn:thf}
\overline{u_i^\prime T^\prime} = - \alpha_t\dx{T}{x_i} 
\end{equation}
The model therefore requires only
the turbulent viscosity provided by the underlying momentum turbulence
model.
In the present work, this correlation is used in combination with the
$k$--$\omega$ SST model as implemented in \ofversion{}
\cite{OpenFOAMv2212}.

\subsubsection{$k_\theta$--$\epsilon_\theta$ model (MM)}
\label{subsubsec:kThetaEpsTheta}

The MM model proposed by \citet{Manservisi2014} is a combined
turbulence model that employs the AKN formulation for momentum
turbulence and a $k_\theta$--$\epsilon_\theta$ closure for thermal
turbulence.
It is a diffusivity-based model, where the turbulent heat flux is modeled with \Cref{eqn:thf}. The corresponding turbulent thermal diffusivity
is given by:
\begin{equation}
\label{eqn:alphatMM}
\alpha_t = C_\theta k \tau_{l\theta} \, 
\end{equation}
where $\tau_{l\theta}$ is the local thermal characteristic time scale, which according to \citet{Manservisi2014} is modeled as
\begin{equation}
\label{eqn:taulthetaMM}
\tau_{l\theta} =
f_{1\theta}\tau_u Pr_{t\infty}
+ f_{2\theta}\tau_m
+ f_{3\theta}\tau_u \sqrt{\frac{2R}{Pr}}
\frac{1.3}{\sqrt{Pr} R_t^{3/4}}\, ,
\end{equation}
in which the following time scales appear:
\begin{equation}
\label{eqn:timescalesMM}
\tau_u = \frac{k}{\epsilon}, \quad
\tau_\theta = \frac{k_\theta}{\epsilon_\theta}, \quad
R = \frac{\tau_\theta}{\tau_u}, \quad
\tau_m = \frac{2 R \tau_u}{R + C_\gamma} 
\end{equation}

The weighting functions $f_{1\theta}$, $f_{2\theta}$ and $f_{3\theta}$
are defined as  in \citet{Manservisi2015}, with corrections and
modifications with respect to the original formulation reported in
\citet{Manservisi2014}:
\begin{align}
\label{eqn:ftheta1}
f_{1\theta} &=
\left( 1 - e^{- R_\delta \sqrt{Pr} / 19} \right)
\left( 1 - e^{- R_\delta / 14} \right) \, , \\
\label{eqn:ftheta2}
f_{2\theta} &= f_{1\theta} \, e^{-(R_t/500)^2} \, , \\
\label{eqn:ftheta3}
f_{3\theta} &= f_{1\theta} \, e^{-(R_t/200)^2} 
\end{align}
The definitions of $R_t$ and $R_\delta$ are given in
\Cref{eqn:Rt_Rdelta}. It should be noted that the expression of
$f_{1\theta}$ differs from the one originally reported in
\citet{Manservisi2014,Manservisi2015}, and follows the corrected form
proposed in subsequent works \cite{Manservisi2016,Manservisi2019}.
In particular, a minus sign is missing in the exponential argument of
the second term in \Cref{eqn:ftheta1} in the original formulation.

The thermal turbulence quantities $k_\theta$ and $\epsilon_\theta$ are
obtained from the following transport equations:
\begin{align}
\label{eqn:kthetaMM}
\frac{\partial k_\theta}{\partial t} + u_i \frac{\partial k_\theta}{\partial x_i} &=
\frac{\partial}{\partial x_i}
\left[
\left( \alpha + \frac{\alpha_t}{\sigma_{k_\theta}} \right)
\frac{\partial k_\theta}{\partial x_i}
\right]
+ P_\theta - \epsilon_\theta \\
\label{eqn:epsilonthetaMM}
\begin{split}
\frac{\partial \epsilon_\theta}{\partial t} + u_i \frac{\partial \epsilon_\theta}{\partial x_i} &=
\frac{\partial}{\partial x_i}
\left[
\left( \alpha + \frac{\alpha_t}{\sigma_{\epsilon_\theta}} \right)
\frac{\partial \epsilon_\theta}{\partial x_i}
\right] + \\
&\quad + \frac{\epsilon_\theta}{k_\theta}
\left( C_{p1} P_\theta - C_{d1} \epsilon_\theta \right)
+ \frac{\epsilon_\theta}{k}
\left( C_{p2} P_k - C_{d2} \epsilon \right) 
\end{split}
\end{align}
The thermal production term  appearing in \Cref{eqn:kthetaMM} is defined as:
\begin{equation}
\label{eqn:Ptheta}
P_\theta = - \alpha_t \frac{\partial T}{\partial x_i}
\frac{\partial T}{\partial x_i} \, ,
\end{equation}
and the model coefficients are selected according to \cite{Manservisi2015}:
\begin{align}
\label{eqn:coefficientsMM}
&C_\theta = 0.1, \quad Pr_{t\infty} = 0.9, \quad C_\gamma = 0.3, \quad
\sigma_{k_\theta} = 1.4, \notag \\
&\sigma_{\epsilon_\theta} = 1.4, \quad
C_{p1} = 0.925, \quad C_{d1} = 1.0, \quad C_{p2} = 0.9
\end{align}
In particular, the coefficient $C_{d2}$ is defined as given in \cite{Manservisi2015}:
\begin{equation}
\label{eqn:Cd2MM}
C_{d2} =
\left[
1.9 \left( 1 - 0.3 e^{-0.0237 R_t^2} \right) - 1
\right]
\left( 1 - e^{-0.1754 R_\delta} \right)^2 
\end{equation}

Different expressions for $C_{d2}$ are reported in the literature.
Specifically, \citet{Manservisi2014} do not include the term $-1$ in
the first bracket, while \citet{Manservisi2015b,Manservisi2016} adopt
the same structure as \Cref{eqn:Cd2MM} but with a different coefficient
in the second exponential (0.0308 instead of 0.1754). In the present work,   
the latter value is adopted, consistently with
the relation proposed by \citet{abe1993}, corresponding to $1/5.7$.

More generally, the formulation of the MM model adopted here is based
on a combination of expressions reported across different publications
by the original authors. In several cases, discrepancies were
identified, including inconsistent coefficients and missing terms. The
final implemented formulation corresponds to a consistent set of
equations obtained through cross-verification of the available sources
and validation against reference results reported in the literature.

\subsubsection{$k_\theta$--$\Omega_\theta$ model (DAVIA)}
\label{subsubsec:kThetaOmegaTheta}

The DAVIA thermal turbulence model is used in combination with the KLW
momentum model (\Cref{subsubsec:keOmega}) and is based on the
formulation proposed by \citet{Manservisi2016}, derived from the
$k_\theta$--$\epsilon_\theta$ model of \citet{Manservisi2014}
(see \Cref{subsubsec:kThetaEpsTheta}).
The model introduces $\Omega_\theta$ as the second thermal turbulence
variable, defined through the transformation $\epsilon_\theta = C_\mu k_\theta e^{\Omega_\theta}$, once plugged into \Cref{eqn:epsilonthetaMM}, yielding:
\begin{align}
\label{eqn:OmegathetaVia}
\frac{\partial \Omega_\theta}{\partial t} + u_i \frac{\partial \Omega_\theta}{\partial x_i}
&=
\frac{\partial}{\partial x_i}
\left[
\left(\alpha + \frac{\alpha_t}{\sigma_{\epsilon_\theta}}\right)
\frac{\partial \Omega_\theta}{\partial x_i}
\right]
+ \frac{2}{k_\theta}
\left( \alpha + \frac{\alpha_t}{\sigma_{\epsilon_\theta}} \right)
\frac{\partial k_\theta}{\partial x_i}
\frac{\partial \Omega_\theta}{\partial x_i} + \notag \\
&\quad
+ \left( \alpha + \frac{\alpha_t}{\sigma_{\epsilon_\theta}} \right)
\frac{\partial \Omega_\theta}{\partial x_i}
\frac{\partial \Omega_\theta}{\partial x_i}
+ \frac{C_{p1}-1}{k_\theta} P_\theta
+ \frac{C_{p2}}{k} P_k + \notag \\
&\quad
- C_\mu (C_{d1} - 1) e^{\Omega_\theta}
- C_\mu C_{d2} e^\Omega 
\end{align}
The production terms $P$ and $P_\theta$ are defined in
\Cref{eqn:prodK,eqn:Ptheta}, respectively. The transport equation for $k_\theta$
remains identical to \Cref{eqn:kthetaMM}.
\Cref{eqn:OmegathetaVia} differs from the formulation reported by
\citet{Manservisi2016} in the coefficient multiplying the term
$(C_{d1}-1)e^{\Omega_\theta}$, where the present implementation
includes the factor $C_\mu$, as required by the previously introduced transformation of $\epsilon_\theta$. 

The thermal time scale $\tau_{l\theta}$ is computed using
\Cref{eqn:taulthetaMM}, consistently with the MM model. This differs
from the formulation originally proposed by \citet{Manservisi2016},
where only the first term is expressed as $f_{1\theta}\tau_u/Pr_{t\infty}$,
i.e. dividing by $Pr_{t\infty}$ instead of multiplying by it, while the
remaining terms are unchanged.

In addition, the value of $Pr_{t\infty}$ is taken equal to $0.9$, as in
the MM model, instead of the value $Pr_{t\infty}=0.75188$ suggested by
\citet{Manservisi2019}. The present implementation therefore differs
from the original DAVIA formulation both in the first term of the time
scale expression and in the value of $Pr_{t\infty}$. 

% \rev{XXX What did yield to the necessity of making changes here? XXX}

The weighting functions $f_{1\theta}$, $f_{2\theta}$ and $f_{3\theta}$
are taken from \citet{Manservisi2016} and correspond to
\Cref{eqn:ftheta1,eqn:ftheta2,eqn:ftheta3}. The coefficient $C_{d2}$ is
defined as in \Cref{eqn:Cd2MM}, following the discussion in
\Cref{subsubsec:kThetaEpsTheta}. The model coefficients are selected according to
\citet{Manservisi2016}:
\begin{align}
C_\mu &= 0.09, \quad
\sigma_{\epsilon_\theta} = 1.4, \quad
C_{p1} = 1.025, \quad
C_{d1} = 1.1 
\end{align}
The coefficient $C_{p2}$ is set to $0.9$, consistently with the MM
model, instead of the value $1.9$ originally proposed by
\citet{Manservisi2016}.

As in the KLW model (\Cref{subsubsec:keOmega}), the cross-diffusion
terms in \Cref{eqn:OmegathetaVia} can be treated either explicitly or
partially implicitly. Both options are available in the present
implementation.

\subsubsection{Algebraic heat-flux model (AHFM)}
\label{subsubsec:ahfm-nrg}

The algebraic heat-flux model considered in this study follows
the formulation proposed by \citet{Shams2014}, based on the model of
\citet{Kenjeres2005}. The adopted formulation corresponds to the
so-called AHFM-NRG model, originally developed for low-$\Pra$-number
flows in forced, natural, and mixed convection regimes
\cite{Shams2014}. 
An improved variant, referred to as AHFM-NRG+, was later introduced by
\citet{Shams2018Number3}. The AHFM-NRG+ modifications concern the buoyancy-production
contribution of the closure (coefficient $C_{t3}$ in \Cref{eqn:AHFMNRG}), which vanishes
identically for $g_i=0$. Consequently, for the purely forced-convection cases
considered here, AHFM-NRG and AHFM-NRG+ coincide, and the latter is not considered
separately.

The AHFM provides a direct algebraic closure for the turbulent heat
flux $\overline{T' u_i'}$, given by:
\begin{align}
\label{eqn:AHFMNRG}
\overline{T'u_i'} &= - C_{t0} \frac{k}{\epsilon}
\left(
C_{t1}\overline{u_i' u_j'} \frac{\partial T}{\partial x_j}
+ C_{t2} \overline{T' u_j'} \frac{\partial u_i}{\partial x_j}
+ C_{t3} \beta g_i \, 2 k_\theta
\right) + \notag \\
&\quad + C_{t4}
\left(
\frac{\overline{u_i' u_j'}}{k} - \frac{2}{3}\delta_{ij}
\right)
\overline{T'u_j'} 
\end{align}
Consistently with the other models, buoyancy effects are neglected in all
simulations ($g_i=0$); the corresponding term is reported here only for
completeness of the published formulation.

The Reynolds stresses $\overline{u_i' u_j'}$, $k$ and $\nu_t$ are provided by the underlying momentum turbulence model, in this case ShamsKE (\cref{subsubsec:shamske}).
The temperature
variance $k_\theta$ is obtained from the following transport equation:
\begin{equation}
\label{eqn:kthetaAHFMNRG}
\frac{\partial (\rho k_\theta)}{\partial t} + u_i \frac{\partial (\rho k_\theta)}{\partial x_i} =
\frac{\partial}{\partial x_i}
\left[
\rho \left(\alpha + \frac{\nu_t}{\sigma_{k_\theta}} \right)
\frac{\partial k_\theta}{\partial x_i}
\right]
+ \rho P_\theta - \rho \epsilon_\theta \, ,
\end{equation}
where $P_\theta$ is defined in \Cref{eqn:Ptheta}. In the AHFM-NRG
formulation, the ratio of turbulent time scales $R$ is assumed
constant, leading to:
\begin{equation}
\epsilon_\theta = \frac{k_\theta}{k} \frac{\epsilon}{R}
\end{equation}
with the constant time-scale ratio $R=0.5$ \citet{Shams2014}. The remaining model
coefficients are $C_{t0}=0.2$, $C_{t2}=0.6$, $C_{t3}=2.5$, $C_{t4}=0.0$ and
$\sigma_{k\theta}=1.0$ \citet{Shams2014}.
It should be noted that \Cref{eqn:kthetaAHFMNRG} differs from
\Cref{eqn:kthetaMM} in the diffusion term, where $\nu_t$ is used
instead of $\alpha_t$.

When the AHFM is used in combination with the ShamsKE model, the
coefficient $C_{t1}$ is defined as \cite{Shams2014}:
\begin{equation}
\label{eqn:AHFM2014}
C_{t1} =
\begin{cases}
0.053 \ln(Re\,Pr) - 0.27, & Re\,Pr > 180 \\
0.25, & Re\,Pr \leq 180
\end{cases} \, 
\end{equation}
A revised expression for $C_{t1}$ was later proposed by
\citet{Shams2019Number3} for use with the EBRSM model, defined as
\[
C_{t1} = 0.176 \ln(Re\,Pr) - 0.426, \quad \text{for } Re\,Pr > 145 
\]
Since no specification is provided for $Re\,Pr \leq 145$ in
\citet{Shams2019Number3}, the value $C_{t1} = 0.25$ is assumed, in
analogy with the original formulation of \citet{Shams2014}
(\Cref{eqn:AHFM2014}). The resulting implementation is therefore:
\begin{equation}
\label{eqn:AHFM2019}
C_{t1} =
\begin{cases}
0.176 \ln(Re\,Pr) - 0.426, & Re\,Pr > 145 \\
0.25, & Re\,Pr \leq 145
\end{cases}
\end{equation}

The present implementation allows switching between the two
formulations of $C_{t1}$. In addition, the algebraic equation
\Cref{eqn:AHFMNRG} can be solved either explicitly or implicitly. In
the explicit approach, the turbulent heat flux appearing on the
right-hand side is taken from the previous iteration, while in the
implicit approach the linear system described by  \Cref{eqn:AHFMNRG} is solved.

In the absence of buoyancy effects, i.e. when the gravitational term
is neglected in \Cref{eqn:AHFMNRG}, the turbulent heat flux becomes
independent of $k_\theta$. Therefore, for purely forced convection
problems, the transport equation for $k_\theta$ does not need to be
solved.

Finally, it should be noted that the coefficient $C_{t1}$ depends on
the Reynolds number $Re$, defined using global reference quantities.
This introduces a non-local dependency, as these quantities are not
uniquely defined in general flow configurations, except for canonical
cases such as channel or pipe flows.

In the present work, $Re$ is taken as the bulk Reynolds number of each
configuration: $Re_b=2\delta\,u_b/\nu$, based on the full channel height,
for the channel; $Re_b=u_b D/\nu$ for the pipe; and the inlet bulk Reynolds
number $Re_b=2\,l_{\mathrm{step}}u_b/\nu$ for the backward-facing step. For
the channel ($Re_\tau=395,640$) and pipe configurations, where $Re\,Pr>180$,
the corresponding $C_{t1}$ value follows directly from \Cref{eqn:AHFM2014}; for
$Re_\tau=180$, where $Re\,Pr\le180$, the clamped value $C_{t1}=0.25$ applies.
For the backward-facing step, the value $C_{t1}=0.0052$ is used, as detailed
in \Cref{subsec:bfs}.

Details on the implementation of the thermal turbulence models within
the OpenFOAM framework, including the class structure and solver
integration, are provided in \Cref{sec:appendix_implementation}. 

% ===== end turbulenceModels.tex =====

\section{Numerical setup and simulation details}
\label{sec:numericalSetup}
% ===== begin numericalSetup.tex =====
This section describes the numerical setup adopted in the simulations.
All cases presented in this study are provided as supplementary
material to ensure full reproducibility and allow for independent
verification and further analysis. In particular, all numerical
settings, solver configurations, and case files can be retrieved on a
case-by-case basis from the supplementary material, which contains the
complete set of simulations performed in this work.

The working fluid is assumed to be Newtonian with constant
thermophysical properties. The effects of gravity and viscous
heating are neglected. The set of governing equations contains the steady-state continuity, momentum, and
energy equations, \Crefrange{eqn:continuity}{eqn:energy}, which in Cartesian
coordinates read as follows:
\begin{align}
\label{eqn:continuity}
& \frac{\partial u_i}{\partial x_i} = 0 \, ,\\
\label{eqn:momentum}
& u_j \frac{\partial u_i}{\partial x_j}
= -\frac{\partial (p/\rho)}{\partial x_i}
+ \nu \frac{\partial^2 u_i}{\partial x_j^2}
- \frac{\partial}{\partial x_j} \left( \overline{u_i' u_j'} \right)
+ S_m \, ,\\
\label{eqn:energy}
& u_i \frac{\partial \theta}{\partial x_i}
= \alpha \frac{\partial^2 \theta}{\partial x_i^2}
- \frac{\partial}{\partial x_i} \left( \overline{u_i' \theta'} \right)
+ S_\theta 
\end{align}

The Reynolds stress tensor $\overline{u_i' u_j'}$ is modeled according
to \Cref{eqn:re_stress}, while the turbulent heat flux
$\overline{u_i' \theta'}$ is modeled using \Cref{eqn:thf} for
diffusivity-based closures and \Cref{eqn:AHFMNRG} for the algebraic
heat-flux model.

In the energy equation, $\theta$ represents the excess temperature for
channel and pipe flows, defined as $\theta = \overline{T}_w - T$, being $\overline{T}_w$ the mean wall temperature. For the
backward-facing step case $\theta$ is defined in \Cref{subsec:bfs}. 

% Since so-called mixed-boundary conditions are considered \cite{Straub2019}, $T_w' = 0$ and thus the turbulent heat flux based on $\theta$ is equivalent to that expressed in terms of temperature fluctuations
% $T'$, i.e. $\overline{u^\prime_i T^\prime} = \overline{u^\prime_i \theta^\prime}$. \textcolor{red}{XXXXX This is not correct? Usually one defines $\theta = \overline{T_w} - T$ so $\overline{u^\prime_i T^\prime} = \overline{u^\prime_i \theta^\prime}$ is true for constant heat flux as well. Problem with the notation?}
% \textcolor{green}{XXX all temperatures are mean temperatures here. Therefore $\theta' = T_w' - T' = T'$ XXX}

For channel and pipe flow simulations, periodic boundary conditions
are applied between the inlet and outlet to reproduce fully developed
conditions. Additional  source terms are thus added to the momentum and energy equations, in order to reproduce the effect of the linear change in pressure (due to friction, i.e. wall momentum flux) and temperature (as required by the chosen non-dimensionalisation) along the streamwise direction in the periodic setting. 

In \Cref{eqn:momentum}, the source term $S_m$ is implemented using the
\texttt{meanVelocityForce} functionality available in \ofversion{}
(\texttt{fvOptions}), enforcing a prescribed bulk velocity through a
spatially uniform forcing term dynamically adjusted during the
simulation.

In \Cref{eqn:energy}, the source term $S_\theta$ is implemented within
a custom solver for the energy equation and takes different
expressions for channel and pipe configurations:
\begin{equation}
\label{eqn:Stheta}
S_\theta =
\begin{cases}
\dfrac{u_1 T^\ast}{\delta} & \text{channel flow} \\
\dfrac{4u_1 T^\ast}{D} & \text{pipe flow}
\end{cases} \, ,
\end{equation}
where $u_1$ denotes the streamwise velocity component and
$T^\ast = q_w/(\rho c_p u_b)$, with $q_w$ the imposed wall heat flux,
$\rho$ the density, $c_p$ the specific heat at constant pressure, and
$u_b$ the bulk velocity.

The spatial discretization employs second-order accurate schemes for
interpolation, as well as for the gradient, divergence, and Laplacian
operators. The cross-diffusion terms arising in the KLW and DAVIA
models (\Cref{subsubsec:keOmega,subsubsec:kThetaOmegaTheta}) are
treated explicitly.

A segregated solution strategy is adopted: the velocity and pressure
fields are solved first, followed by the temperature field. This
approach is justified by the assumption of constant fluid properties,
under which the momentum equations are decoupled from the thermal
field. Following OpenFOAM terminology, this strategy is referred to as
a \texttt{frozen-flow} approach.
The SIMPLE algorithm is used for the pressure-velocity coupling.

Under-relaxation factors are applied to all equations. Simulations are
initialized with conservative values (typically 0.1), which are
gradually increased to standard values during the iterative process to enhance
convergence. Initial conditions are prescribed as uniform fields with
values of the same order of magnitude as the expected solution.

Convergence to steady state is assumed when the normalized residuals
of all momentum and thermal turbulence quantities fall below
$10^{-9}$ in all reported cases. 

\subsection{Boundary conditions}
\label{subsec:BC}

Mixed boundary conditions are applied at the wall for the thermal
turbulence variables, following \citet{Straub2019}:
\begin{align}
\label{eqn:kThetaWall}
 \left. k_\theta \right|_w &= 0 \, ,\\
\label{eqn:epsilonThetaWall}
 \left.\epsilon_\theta\right|_w &= \alpha \frac{2 k_\theta}{d_w^2} \, ,\\
\label{eqn:OmegaThetaWall}
 \left.\Omega_\theta\right|_w &= \ln\left( \frac{2 \alpha}{C_\mu d_w^2} \right) \, ,\\
\label{eqn:OmegaWall}
 \left.\Omega\right|_w &= \ln\left( \frac{2 \nu}{C_\mu d_w^2} \right) 
\end{align}
where $d_w$ is the distance to the nearest wall, $k_\theta$ in \Cref{eqn:epsilonThetaWall} is evaluated at the cell center, $C_\mu = 0.09$ and  the subscript $w$ denotes the wall. 
The condition $k_\theta|_w=0$ corresponds to an ideally isothermal wall. For
the isoflux conditions considered here, temperature fluctuations at the wall
do not vanish in general. However, all assessed closures were calibrated
with $k_\theta|_w=0$, and the reference data of \citet{Straub2019} show that,
at the Prandtl numbers of interest, the influence of the thermal wall
boundary condition on the mean temperature is small, while it mainly
affects the near-wall temperature variance. This aspect must be kept in
mind when comparing the $\sqrt{2k_\theta^{+}}$ profiles with the reference
data.

% The boundary conditions for further variables are summarized in Table 2.
% It should be noted that the elliptic relaxation factor $f$ and the
% Reynolds stress components $R_{ij}$ are required and thus solved only for
% the EBRSM model. For $\varepsilon$, $k$, and $\omega$, the asymptotic
% near-wall expressions reported in Table 2 are used.
% Boundary conditions for further variables are summarized in
% \Cref{tab:BCs}. It should be noted that the elliptic relaxation factor $f$ and the Reynolds stress components
% $R_{ij}$ are required and thus solved only for the EBRSM model. For $\epsilon$, $k$, and
% $\omega$, the standard low-$\Rey$ wall functions implemented in
% \ofversion{} \cite{OpenFOAMv2212} are used.

%\textcolor{red}{XXX here it is not clear what $k_\theta$ in \cref{eqn:epsilonThetaWall} is. It can't be $\left. k_\theta \right|_w$ because then you would just write 0}
%\textcolor{green}{XXX description added XXX}

\begin{table}[!htbp]
\centering
\caption{Boundary conditions at the wall (values reported without units)}
\begin{tabularx}{0.7\textwidth}{X X}
\toprule
\textbf{Variable} & \textbf{Boundary condition} \\
\midrule
$\epsilon$        & $2\nu k_{cc}/d_w^{2}$ \\
$f$               & 0 \\
$k$               & 0 \\
$\Omega$          & \Cref{eqn:OmegaWall} \\
$\omega$          & $6\nu/(\beta_1 d_w^{2})$ \\
$p$               & zero-gradient \\
$R_{ij}$          & $10^{-20}$ \\
$u_i$             & 0 \\
$\epsilon_\theta$ & \Cref{eqn:epsilonThetaWall} \\
$k_\theta$        & $10^{-20}$ \\
$\Omega_\theta$   & \Cref{eqn:OmegaThetaWall} \\
$\theta$          & 0 \\
\bottomrule
\end{tabularx}
\label{tab:BCs}
\end{table}
The boundary conditions for further variables are summarized in Table 2.
Here, $k_{cc}$ denotes the turbulent kinetic energy at the centre of the
near-wall adjacent cell, and $\beta_1=0.075$ is the standard inner-blending
coefficient of the k--$\omega$ SST model \cite{Menter2003}. 
The expressions for $k$,
$\varepsilon$ and $\omega$ correspond to those selectable through the
respective low-Reynolds wall functions (\texttt{kLowReWallFunction},
\texttt{epsilonWallFunction}, \texttt{omegaWallFunction}) in \ofversion{} \cite{OpenFOAMv2212}.
It should be noted that the
elliptic relaxation factor $f$ and the Reynolds stress components $R_{ij}$
are required and thus solved only for the EBRSM model.
The values $10^{-20}$ prescribed for $k_\theta$ and $R_{ij}$ are numerically equivalent to zero and are used instead of an exact zero for solver stability.
% ===== end numericalSetup.tex =====
   
\section{Results and discussion}
\label{sec:resultsAndDiscussion}
% ===== begin resultsAndDiscussion.tex =====
The following investigations assess the turbulence models in channel
flow, pipe flow, and backward-facing step configurations with three
objectives. First, to verify the reproducibility of results reported in
the literature, given the discrepancies between the original model
formulations and their implementation in the present work
(see \Cref{subsec:model_selection}). Second, to evaluate the numerical
robustness and convergence behavior of the models. Third, to assess
their predictive accuracy against high-fidelity reference data for
low-Prandtl-number flows ($Pr \ll 1$).

For the \kwsst{} model, no additional verification of reproducibility
is required, as the model is adopted without modification from
\citet{OpenFOAMv2212} and has been extensively validated in the
literature.

\subsection{Channel Flow}
\label{subsec:channel}
A quasi-one-dimensional setup is adopted, with a single computational cell in both the streamwise and spanwise
directions, and 200 cells in the wall-normal direction.
A grid stretching is applied in the wall-normal direction with an expansion ratio of 10 from the wall towards the channel centreline. This ensures that the viscous sublayer is properly resolved in all simulations, with \(y^+ < 1\), allowing for the use of a wall resolved approach (sometimes also called low-Reynolds-number in OpenFOAM).
A constant wall heat flux is imposed at both walls. By varying the bulk velocity \(u_b\), simulations are performed at friction Reynolds numbers
\(Re_\tau = u_\tau\delta/\nu = 180,\,395,\,640\), where \(u_\tau = \sqrt{\tau_w/\rho}\) is the friction velocity and \(\delta\) is the channel's half height. A Prandtl number of \(Pr = 0.025\) is
considered. The results are primarily presented for \(Re_\tau = 395\), as similar trends are observed for the other Reynolds numbers.

The DNS data used for comparison with the RANS simulations originates from the DNS database provided by \citet{KawamuraDataBase}. 
The results presented in viscous units are scaled with $u_\tau$, $T_\tau = q_w/(\rho\,c_p\,u_\tau)$ and $\nu$.
 
\begin{figure}[ht]
    \centering
        \begin{subfigure}{0.48\textwidth}
        \centering
        \includegraphics[width=\linewidth]{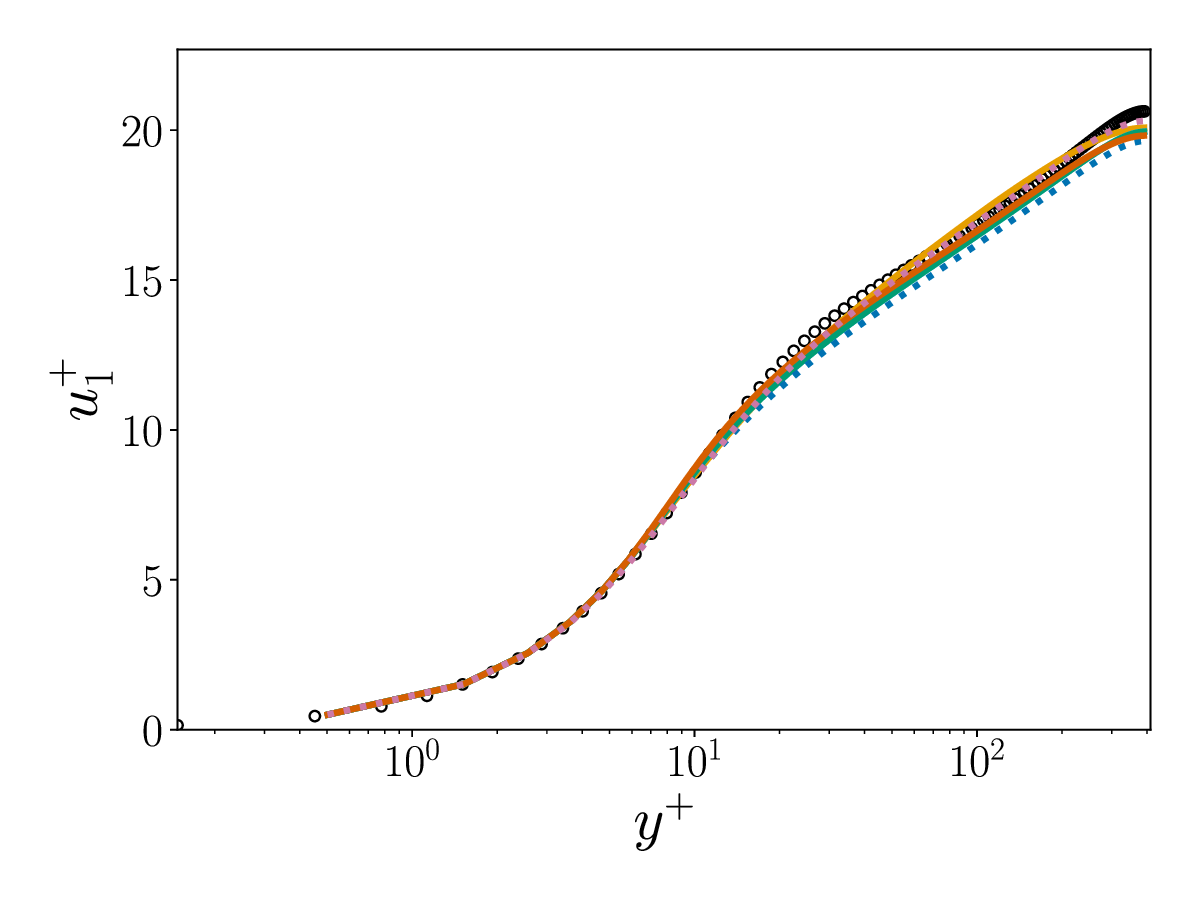}
        \caption{}
        \label{subfig:CF395_u}
    \end{subfigure}
    \hfill
    \begin{subfigure}{0.48\textwidth}
        \centering
        \includegraphics[width=\linewidth]{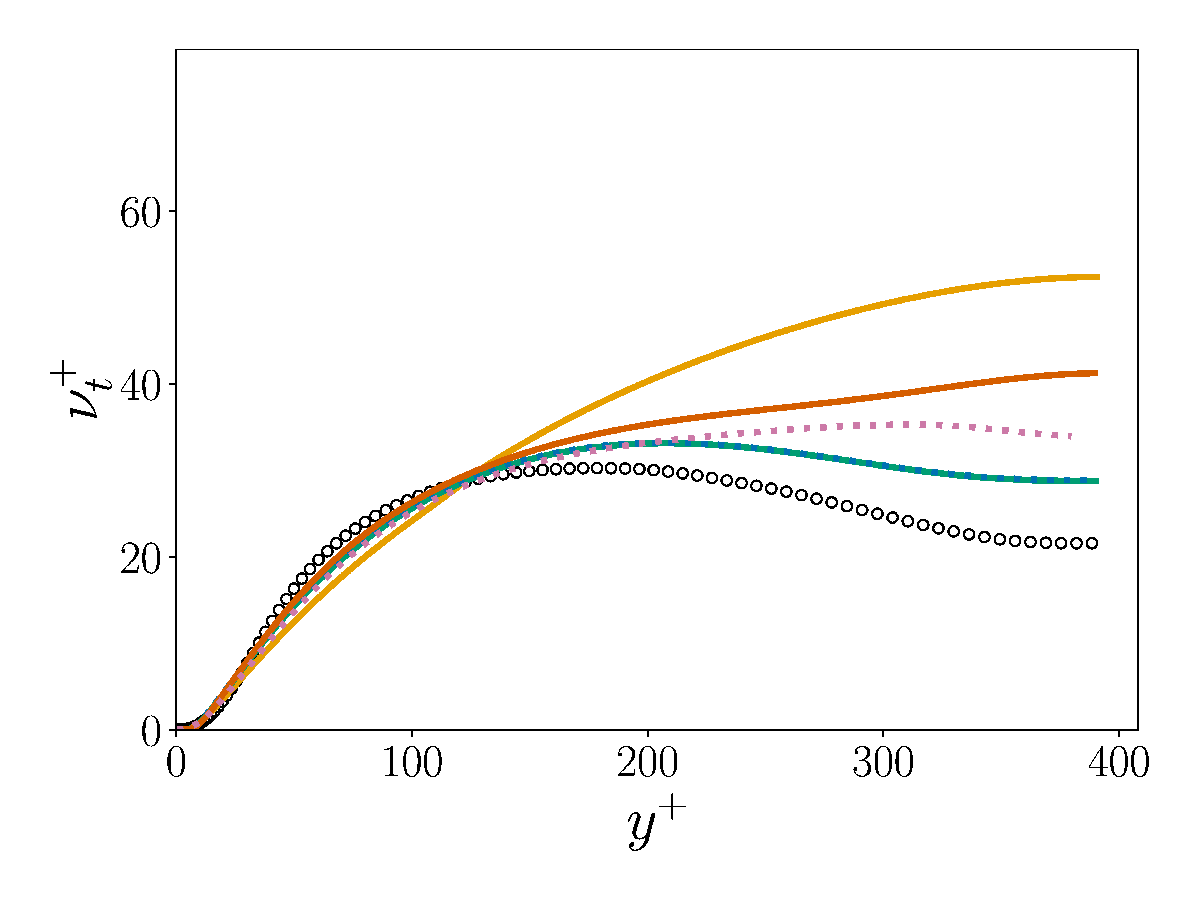}
        \caption{}
        \label{subfig:CF395_nut}
    \end{subfigure}
    % Legend included after last figure
    \renewcommand{\arraystretch}{1.1}\tiny
    \begin{tabular}{|>{\centering\arraybackslash}p{0.1cm}>{\centering\arraybackslash}p{1cm}|>{\centering\arraybackslash}p{0.1cm}>{\centering\arraybackslash}p{1cm}|>{\centering\arraybackslash}p{0.1cm}>{\centering\arraybackslash}p{1cm}|}
        \hline
        \textcolor[HTML]{E69F00}{\rule[0.5ex]{0.3cm}{1.0pt}} & KWSST &  \textcolor[HTML]{0072B2}{\rule[0.5ex]{0.05cm}{1.0pt}}\textcolor[HTML]{FFFFFF}{\rule[0.5ex]{0.05cm}{1.0pt}}\textcolor[HTML]{0072B2}{\rule[0.5ex]{0.05cm}{1.0pt}}\textcolor[HTML]{FFFFFF}{\rule[0.5ex]{0.05cm}{1.0pt}}\textcolor[HTML]{0072B2}{\rule[0.5ex]{0.05cm}{1.0pt}} & KLW & \textcolor[HTML]{CC79A7}{\rule[0.5ex]{0.05cm}{1.0pt}}\textcolor[HTML]{FFFFFF}{\rule[0.5ex]{0.05cm}{1.0pt}}\textcolor[HTML]{CC79A7}{\rule[0.5ex]{0.05cm}{1.0pt}}\textcolor[HTML]{FFFFFF}{\rule[0.5ex]{0.05cm}{1.0pt}}\textcolor[HTML]{CC79A7}{\rule[0.5ex]{0.05cm}{1.0pt}} & EBRSM  \\
        \hline
        \textcolor[HTML]{009E73}{ \rule[0.5ex]{0.3cm}{1.0pt}} & AKN & \textcolor[HTML]{D55E00}{\rule[0.5ex]{0.3cm}{1.0pt}} & ShamsKE   & \simFalseSmall & DNS\\
        \hline
    \end{tabular}
    \caption{(a) streamwise velocity $u_1^+ = u_1/u_\tau$ (b) Turbulent viscosity $\nu_t^+ = \nu_t / \nu$.  DNS data from \cite{KawamuraDataBase}.}
\label{fig:CF395_u_nut}
\end{figure}

As shown in \Cref{subfig:CF395_u}, all turbulence models accurately reproduce the velocity profile $u^+_1$ in viscous units for $Re_\tau=395$. Minor deviations are observed for the KLW and ShamsKE models,
both of which underestimate the velocity in the channel center
compared to their respective reference studies.

The comparison of turbulent viscosity is shown in \Cref{subfig:CF395_nut}. The AKN and KLW models provide the best agreement with the DNS profile, with nearly identical curves. This should not surprise, since  the KLW momentum turbulence model is
derived from the AKN formulation. In contrast, for the EBRSM, which does not employ an eddy viscosity in its
momentum closure, $\nu_t$ is evaluated a posteriori as the ratio of the turbulent
shear stress to the mean strain rate,
$\nu_{t,\mathrm{eff}}=-\overline{u_1'u_2'}/(\partial u_1/\partial x_2)$,
consistently with the evaluation of the DNS data. The resulting profile is in
agreement with the DNS data in the near-wall region, while a moderate overestimation
is observed towards the channel centre, where both the turbulent shear stress and
the mean velocity gradient vanish and their ratio becomes increasingly sensitive to
small differences between the two quantities.

\begin{figure}[ht]
    \centering
    \includegraphics[width=8cm, height=5cm]{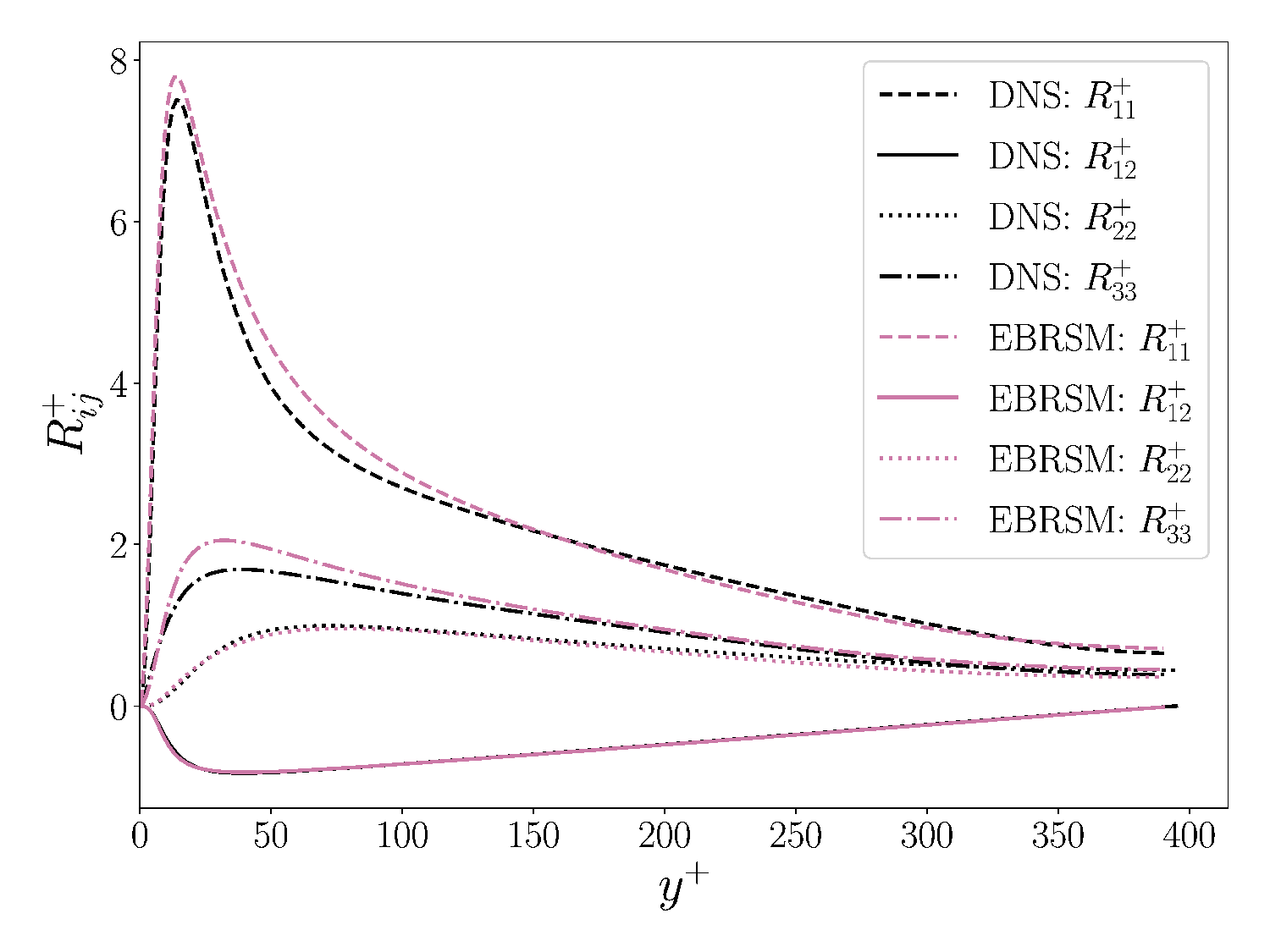}
    \caption{Reynolds stresses in wall units ($R^+_{ij}=\overline{u_i^\prime u_j^\prime}/u_\tau^2$) for EBRSM. DNS data from \cite{KawamuraDataBase}. Channel flow at $Re_\tau = 395$.}
    \label{fig:CF395_Rij}
\end{figure}

A complementary assessment of the EBRSM model is provided by the direct comparison of its Reynolds stress components with the DNS data, as shown in \Cref{fig:CF395_Rij}. 
%
% Since \citet{Shams2019Number3} did not specify the EBRSM model coefficients used, these values were determined through systematic parameter testing. By setting $A_1 = 0.1$ and $C_l = 0.122$, the Reynolds stress results
% reported by \citet{Shams2019Number3} could be reproduced. 
%
The deviations between the EBRSM model results and the DNS data for the Reynolds stresses are minimal.

No significant numerical stability issues were observed for the
KWSST-KAYS and AKN-MM combinations. Small variations in initial conditions did not lead to divergence, and convergent solutions could be obtained with equation relaxation factors above 0.7. The ShamsKE-AHFM model, when used without the Yap correction, also
exhibited robust behavior with respect to the sensitivity to initial
conditions. However, the Yap correction needed to be introduced gradually during the simulation. If the Yap correction was activated from the start with uniform initial conditions, no convergence could be achieved.

In contrast, the KLW-DAVIA and EBRSM-AHFM combinations required
significantly smaller equation relaxation factors and were more
sensitive to initial conditions. Increasing the initial values of $\Omega$ and $R_{ij}$ by a factor of 10 resulted in solution divergence for both models. Thus, the improved numerical robustness claimed by \citet{Manservisi2016} for transitioning from the $k$-$\epsilon$ to the $k$-$\Omega$ model was not observed.
This behaviour is consistent with the logarithmic nature of
$\Omega=\ln\omega$. In the robustness test reported here,
the perturbation was applied directly to the logarithmic variable ($\Omega$
increased by a factor of 10), rather than to the physical dissipation rate
$\omega$. Since $\Omega=\ln\omega$, multiplying $\Omega$ by a factor of 10
corresponds to raising the physical variable to the tenth power,
$\omega_{\mathrm{new}}=(\omega_{\mathrm{old}})^{10}$, rather than to a
comparably moderate rescaling of $\omega$ itself. This is a substantially more
severe perturbation than the initial-condition variations considered by
\citet{Manservisi2016}, and directly explains the divergence observed here
through the exponential source terms $e^{\Omega}$ and
$e^{\Omega_\theta}$, without contradicting
the robustness reported therein for physically-scaled initial fields.
%\textcolor{red}{XXX Does it make sense to put convergence discussion in between the momentum and temperature results?}

In principle, the solution of the momentum equation, represented in nondimensional form, should depend only on the Reynolds number and
not on the specific values of the dimensional quantities used to define
it. However, for the EBRSM model, a different behaviour was observed. When using a low kinematic viscosity (of order \(10^{-7}\,\mathrm{m^2/s}\)), no converged turbulent solution could be obtained, and the simulations invariably relaminarized, even at \(Re_\tau = 395\). In contrast, using a higher viscosity (of order
\(10^{-3}\,\mathrm{m^2/s}\)) at the same Reynolds number led to a
converged turbulent solution. Since the
model equations are dimensionally consistent and the non-dimensional
solution depends on the Reynolds number only, this behaviour must be
attributed to the numerical treatment rather than to the model
formulation. Possible mechanisms include absolute solver tolerances and
dimensional clipping and limiting operations in the implementation, whose
effect does not scale with the magnitude of the flow variables.
% Consequently, all SHAMS-EBRSM results
% presented here are obtained using the latter order of magnitude of \(\nu\). 

\begin{figure}[ht]
    \centering
    \begin{subfigure}{0.48\textwidth}
        \centering
        \includegraphics[width=\linewidth]{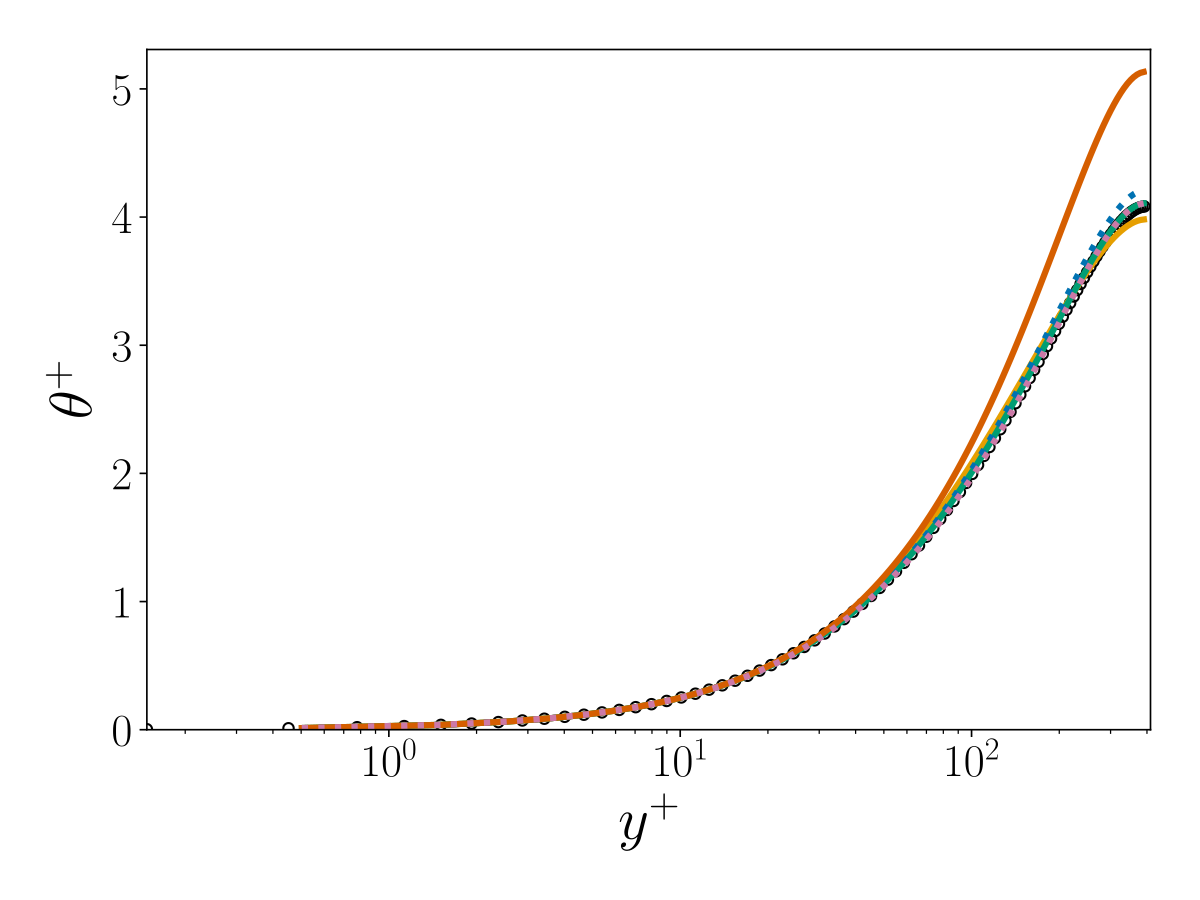}
        \caption{}
        \label{subfig:CF395_T}
    \end{subfigure}
    \hfill
    \begin{subfigure}{0.48\textwidth}
        \centering
        \includegraphics[width=\linewidth]{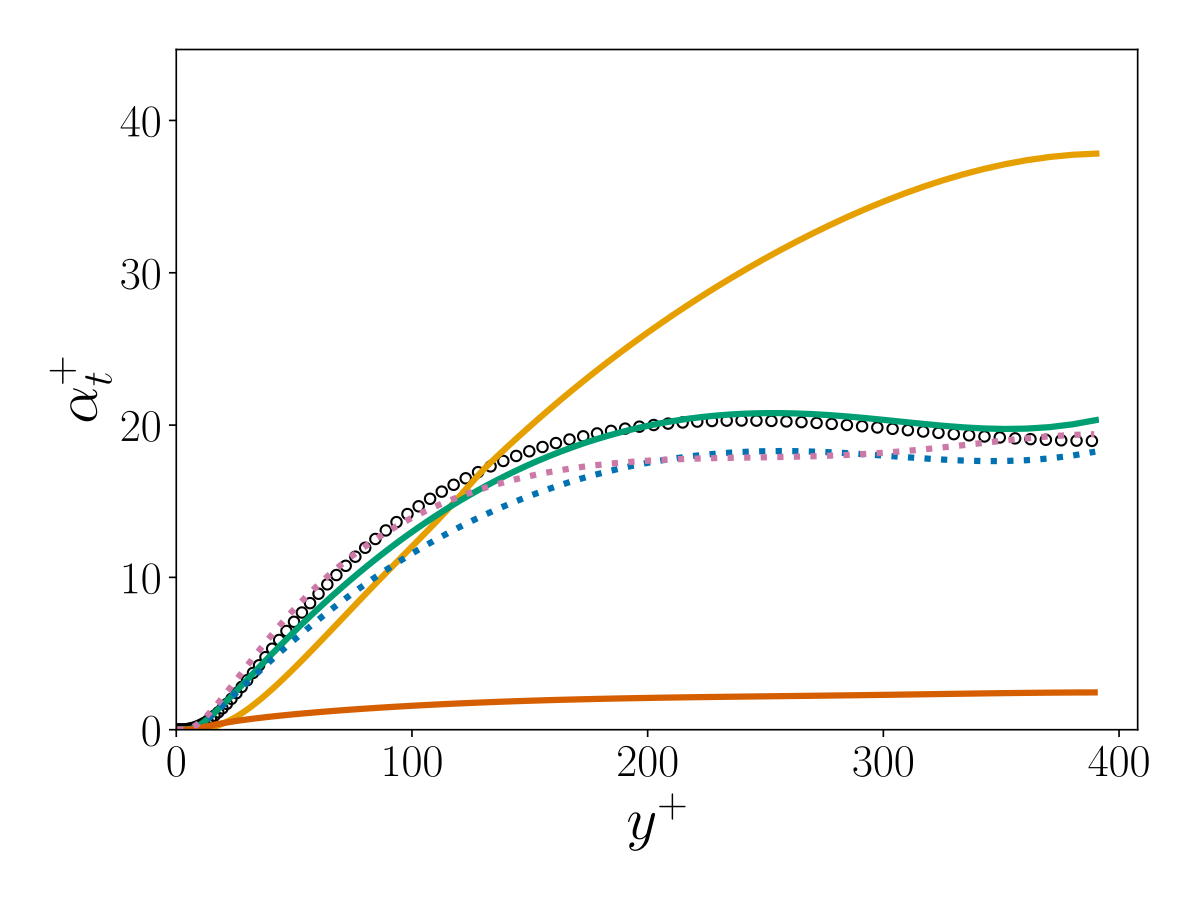}
        \caption{}
        \label{subfig:CF395_alphat}
    \end{subfigure}
    
    \begin{subfigure}{0.48\textwidth}
        \centering
        \includegraphics[width=\linewidth]{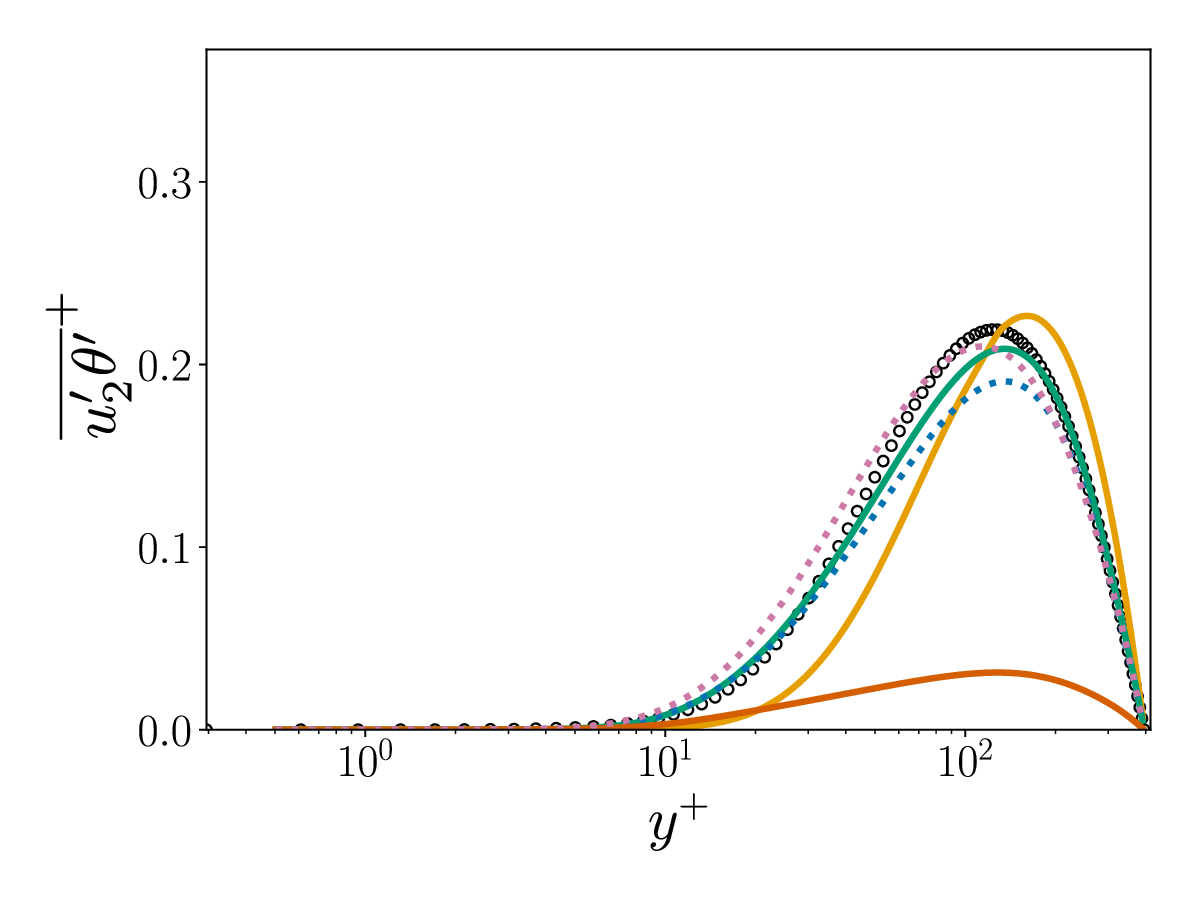}
        \caption{}
        \label{subfig:CF395_u2T}
    \end{subfigure}
    \hfill
    \begin{subfigure}{0.48\textwidth}
        \centering
        \includegraphics[width=\linewidth]{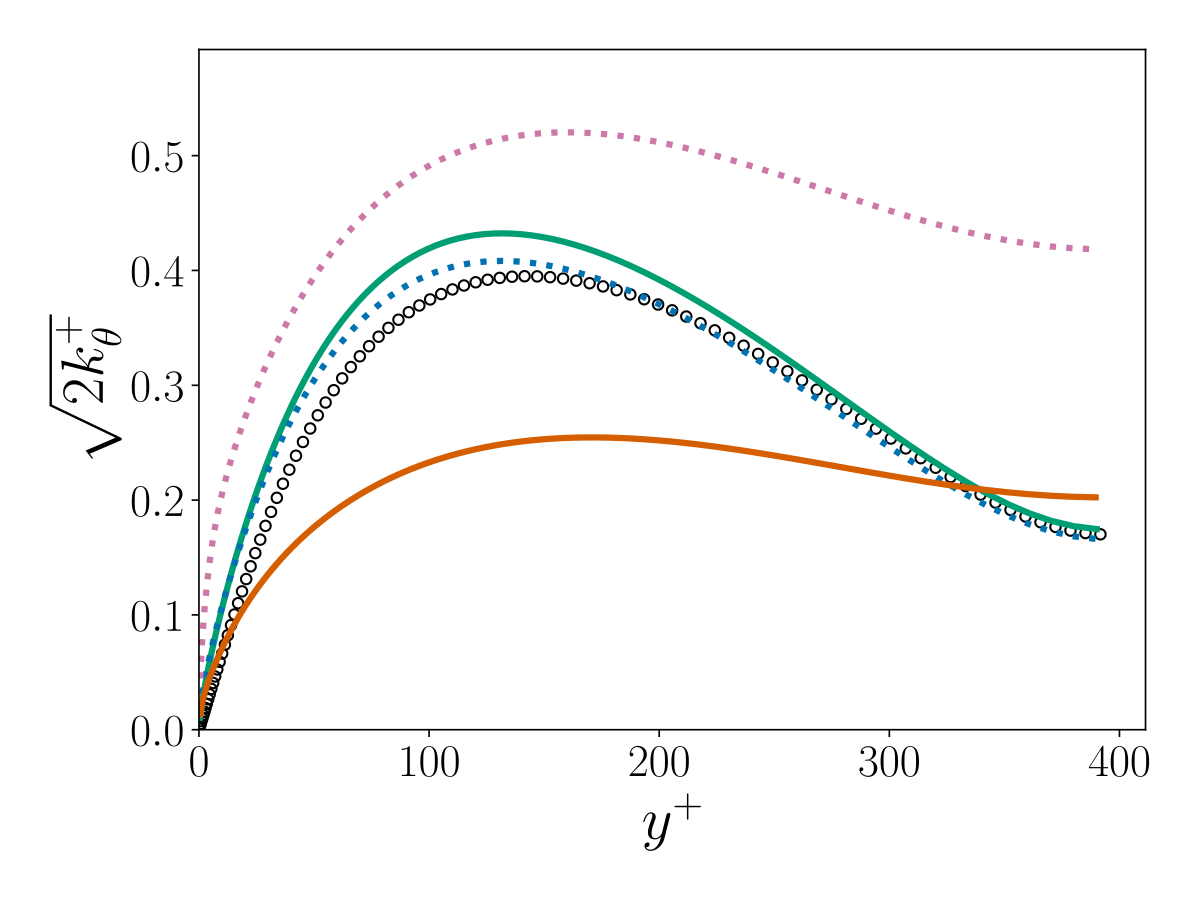}
        \caption{}
        \label{subfig:CF395_kTheta}
    \end{subfigure}\vspace{0.5cm}
    % Legend included in first figure
\renewcommand{\arraystretch}{1.1}\tiny
\begin{tabular}{|>{\centering\arraybackslash}p{0.1cm}>{\centering\arraybackslash}p{2cm}|
                >{\centering\arraybackslash}p{0.1cm}>{\centering\arraybackslash}p{2cm}|}
    \hline
    \textcolor[HTML]{E69F00}{\rule[0.5ex]{0.3cm}{1.0pt}} & KWSST-KAYS &
    \textcolor[HTML]{009E73}{\rule[0.5ex]{0.3cm}{1.0pt}} & AKN-MM \\
    \hline
    \textcolor[HTML]{D55E00}{\rule[0.5ex]{0.3cm}{1.0pt}} & ShamsKE-AHFM &
    \simFalseSmall & DNS \\
    \hline
    \tikz[baseline=-0.5ex]{\draw[colKLW, dotted, line width=1.2pt] (0,0) -- (0.3,0);} & KLW-DAVIA &
    \tikz[baseline=-0.5ex]{\draw[colEBRSM, dotted, line width=1.2pt] (0,0) -- (0.3,0);} & EBRSM-AHFM \\
    \hline
\end{tabular}
    \caption{(a) normalized temperature $\theta^+ =\theta/T_\tau$, (b) turbulent thermal diffusivity
(\(\alpha_t^+ = \alpha_t / \alpha\)), (c) wall-normal turbulent
heat flux $\overline{u_2^\prime\theta^\prime}^+ = \overline{u_2^\prime \theta^\prime}/(u_\tau T_\tau)$ (d) temperature standard deviation
$\sqrt{2k_\theta^+} = \sqrt{2k_\theta}/T_\tau$. DNS data from \citet{KawamuraDataBase}.}
    \label{fig:CF395_T_alphat_vT_kTheta}
\end{figure}

% Thermal results
The results for the temperature $\theta$, turbulent thermal diffusivity $\alpha_t$, and turbulent heat flux component normal to the wall $\overline{u_2'\theta'}$ in wall units are shown in \Cref{fig:CF395_T_alphat_vT_kTheta}. For the two AHFM-based combinations (ShamsKE-AHFM and EBRSM-AHFM), which do not rely on the
gradient-diffusion hypothesis, the thermal diffusivity is computed a
posteriori as \(\alpha_t = - \overline{u_2^\prime \theta^\prime}/(\partial \theta/\partial x_2)\), where the  subscript $2$ denotes the wall-normal direction. 

The KWSST-KAYS combination closely reproduces the DNS temperature profile,
although it underestimates the wall-normal turbulent heat flux (\Cref{subfig:CF395_u2T}). At
$Pr=0.025$, molecular conduction ($\alpha^{+}=1/Pr=40$) remains the dominant transport
mechanism over most of the channel, so that the mean temperature is only weakly
sensitive to inaccuracies in $\alpha_t$. The accurate temperature prediction must
therefore be attributed primarily to the very low Prandtl number.

Note that, at the low Prandtl number considered here, bulk metrics such as
the Nusselt number are only weakly sensitive to inaccuracies in the modeled turbulent
heat flux, as molecular conduction dominates the energy balance. The profile-resolved comparisons of temperature and turbulent heat flux
presented here are therefore more diagnostic of the underlying closure behaviour than
an integrated heat-transfer coefficient would be. This does not apply to the local
Nusselt number reported for the backward-facing step (\Cref{subsec:bfs}, \Cref{subfig:BFS_Nu}), where
convective transport in the recirculation and reattachment regions, rather than
molecular conduction, controls the local heat transfer.

The AKN-MM, KLW-DAVIA, and EBRSM-AHFM combinations successfully
reproduce the
results reported by their respective authors in the literature.
However, the ShamsKE-AHFM model does not match the results presented by
\citet{Shams2014} and shows significant deviations from DNS data at
\(Re_\tau = 395\), with similar behaviour also observed at other
Reynolds numbers. Due to this inconsistency, the ShamsKE-AHFM combination is excluded from
the accuracy assessment. Its results are nevertheless reported for the
pipe and backward-facing step configurations in order to verify whether
the lack of reproducibility is systematic and thereby to isolate its
origin.
It should be noted that different implementations of the ShamsKE-AHFM
model exist across CFD platforms. In particular, \citet{Shams2014} performed their simulations using STAR-CCM+ \cite{STARCCMv704}, while \citet{Shams2019Number3} employed CODE SATURNE \cite{CodeSaturne}, whereas the present work is based on
\ofversion. Due to the limited level of detail provided in the original
publications, these implementations may differ in terms of model
coefficients, auxiliary functions, and numerical treatment. 
This lack of a uniquely defined and fully documented formulation can
lead to significant discrepancies between reported results and
independent implementations. The deviations observed in the present work,
particularly for the ShamsKE model, are therefore likely not only
related to numerical aspects, but also to differences in the underlying
model formulation.

Since the results obtained with the ShamsKE momentum model are able to
reproduce those reported by \citet{Shams2014}, it can be inferred that
the observed discrepancies are likely associated with the implementation
or modeling of the explicit expression for the turbulent heat flux,
rather than with the underlying momentum turbulence model.
The fact that the EBRSM-AHFM combination is able to
reproduce the results reported by \citet{Shams2019Number3}, obtained
with a different CFD code, does not allow isolating the exact origin of
the discrepancies observed for the ShamsKE-AHFM combination. 

The numerical robustness issues observed for the EBRSM and KLW models in the channel flow configuration could, in principle, justify their
exclusion from further analysis. However, with appropriate numerical settings, both models are able to reproduce the results reported by their respective authors and show good agreement with DNS data.
For this reason, they are retained and further evaluated in turbulent pipe flow simulations, as discussed in \Cref{subsec:pipe}.

\FloatBarrier

\subsection{Pipe Flow}
\label{subsec:pipe}
\begin{figure}[ht]
    \centering
    \includegraphics[width=5cm, height=5cm]{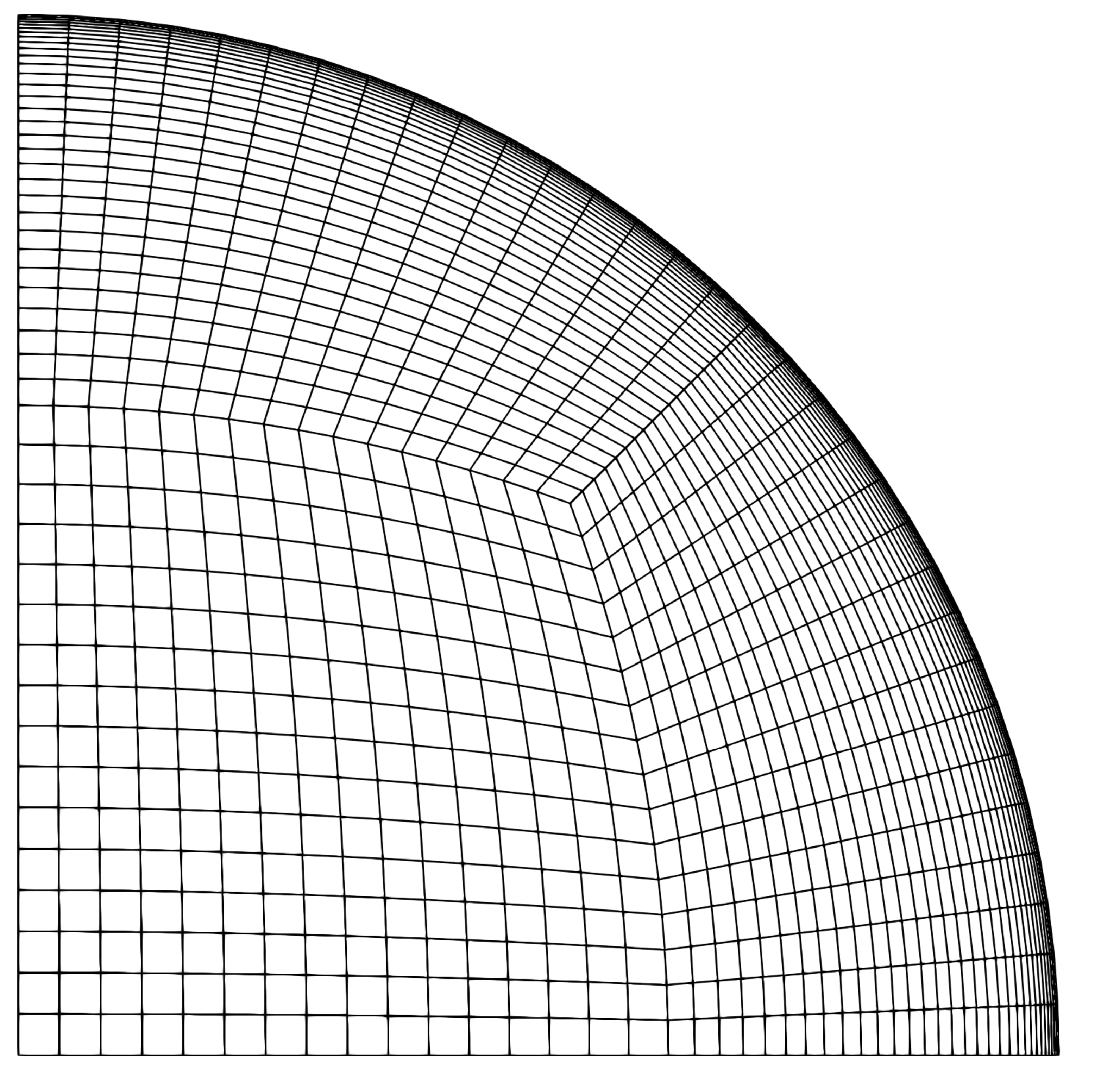}
    \caption{One-fourth of the mesh in the pipe cross section}
    \label{fig:PF-mesh}
\end{figure}

The turbulent pipe flow configuration is considered as a second
canonical test case to further assess the reproducibility of the
different turbulence models beyond the channel flow.
A fully developed flow is simulated using periodic boundary conditions
in the streamwise direction. A constant and spatially uniform wall heat
flux is imposed along the pipe wall.
The simulations are performed on a so-called \textit{O-ring mesh},
shown in \Cref{fig:PF-mesh}, to avoid singularities at the pipe centerline,
while ensuring sufficient near-wall resolution to capture the viscous
sublayer. A single cell is used in the streamwise direction, while 4992
cells are employed in the pipe cross-section.

The same modeling assumptions as in the channel flow are retained,
including constant fluid properties, neglect of buoyancy effects, and
the use of a segregated solution strategy in which the flow field is
solved prior to the thermal field (\texttt{frozen-flow} approach).
Compared to the channel flow, the pipe configuration introduces an
increased geometrical complexity due to the non Cartesian, not exactly orthogonal  mesh, allowing
for a more stringent assessment of numerical robustness.

The data used for comparison with the RANS simulations originate from
the database provided by \citet{StraubDataBase}. No pipe flow simulations were presented for the EBRSM-AHFM combination in
\citet{Shams2019Number3}.
The simulations are performed at a bulk Reynolds number
$Re_b = u_bD/\nu = 11700$ and a Prandtl number $Pr=0.025$, matching the
conditions of the reference database, where $u_b$ is the imposed bulk
velocity and $D$ is the pipe diameter.

\begin{figure}[ht]
    \centering
    \begin{subfigure}{0.48\textwidth}
        \centering
        \includegraphics[width=\linewidth]{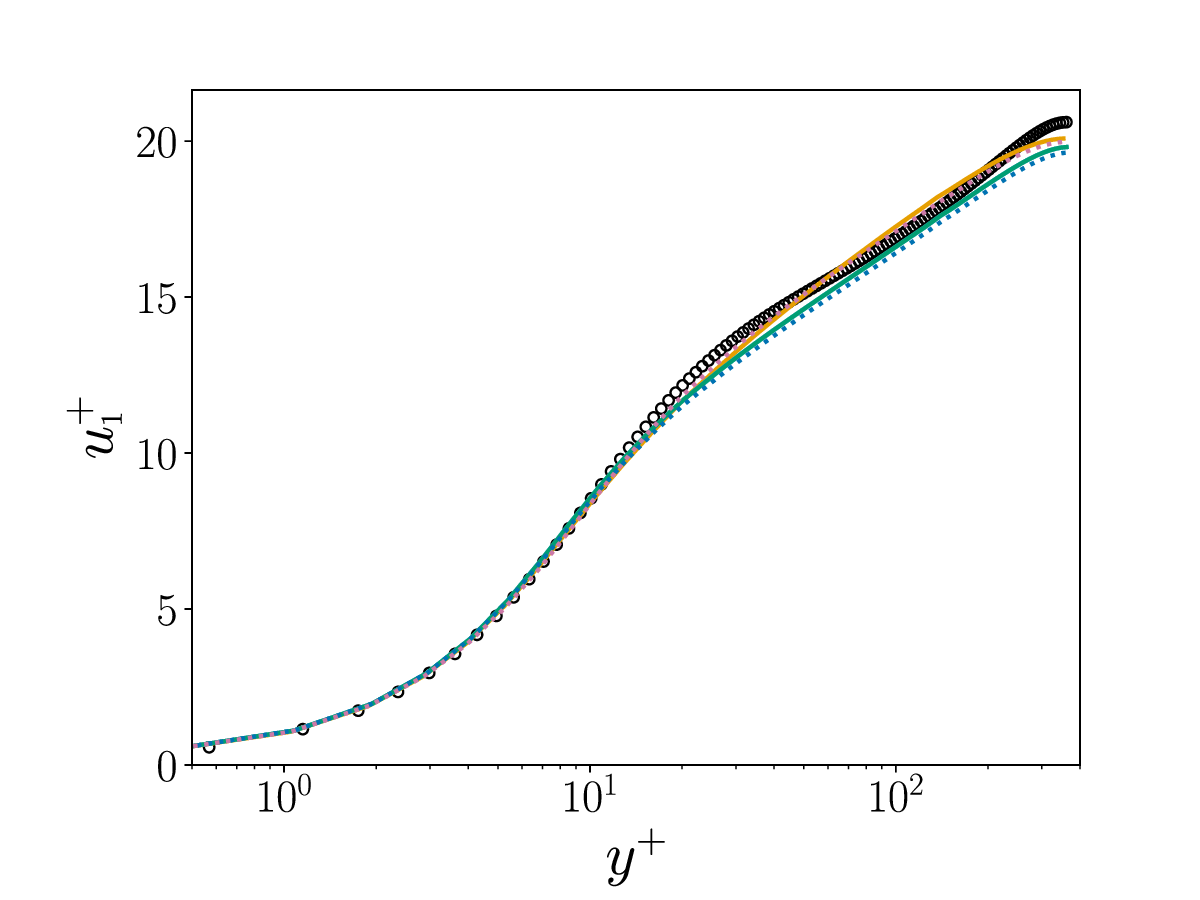}
        \caption{}
        \label{subfig:PF_u}
    \end{subfigure}
    \hfill
    \begin{subfigure}{0.48\textwidth}
        \centering
        \includegraphics[width=\linewidth]{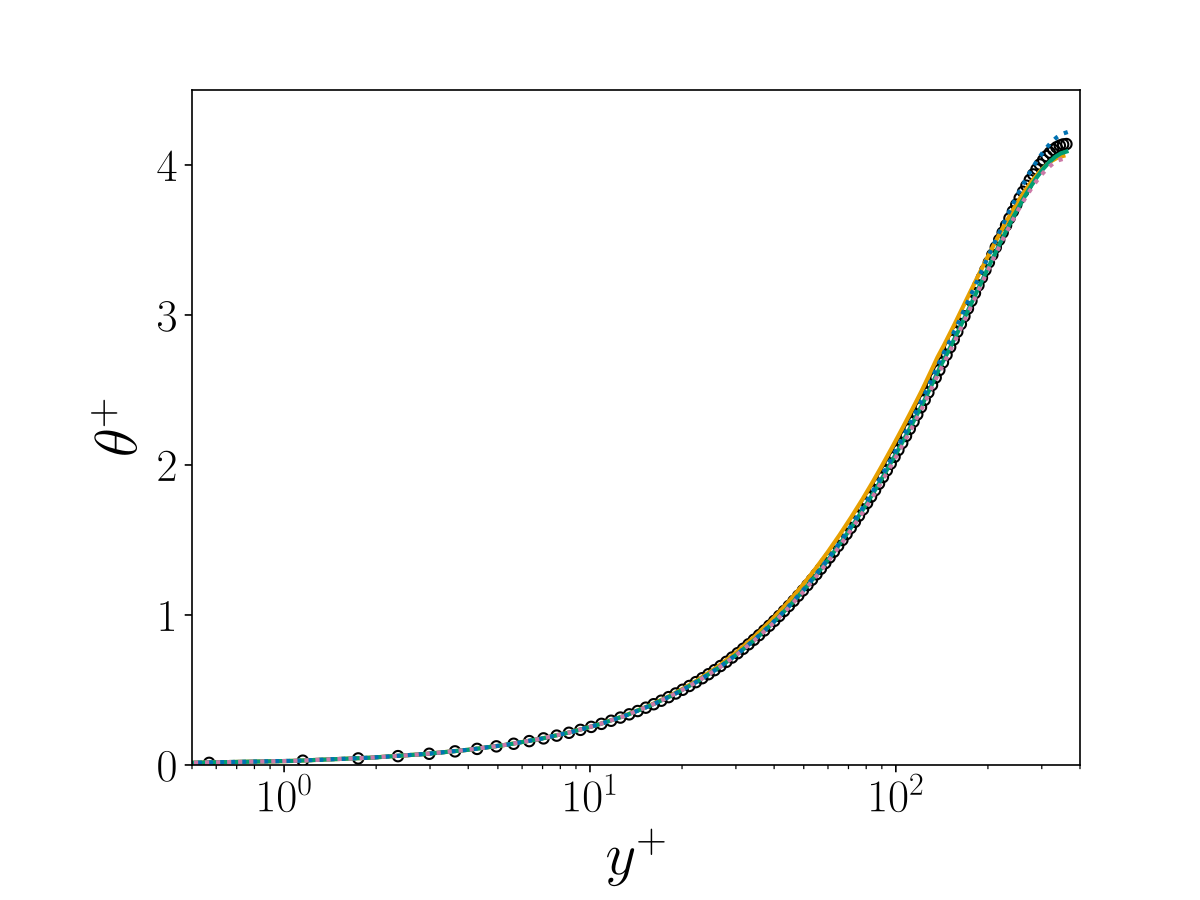}
        \caption{}
        \label{subfig:PF_T}
    \end{subfigure}
    
    \begin{subfigure}{0.48\textwidth}
        \centering
        \includegraphics[width=\linewidth]{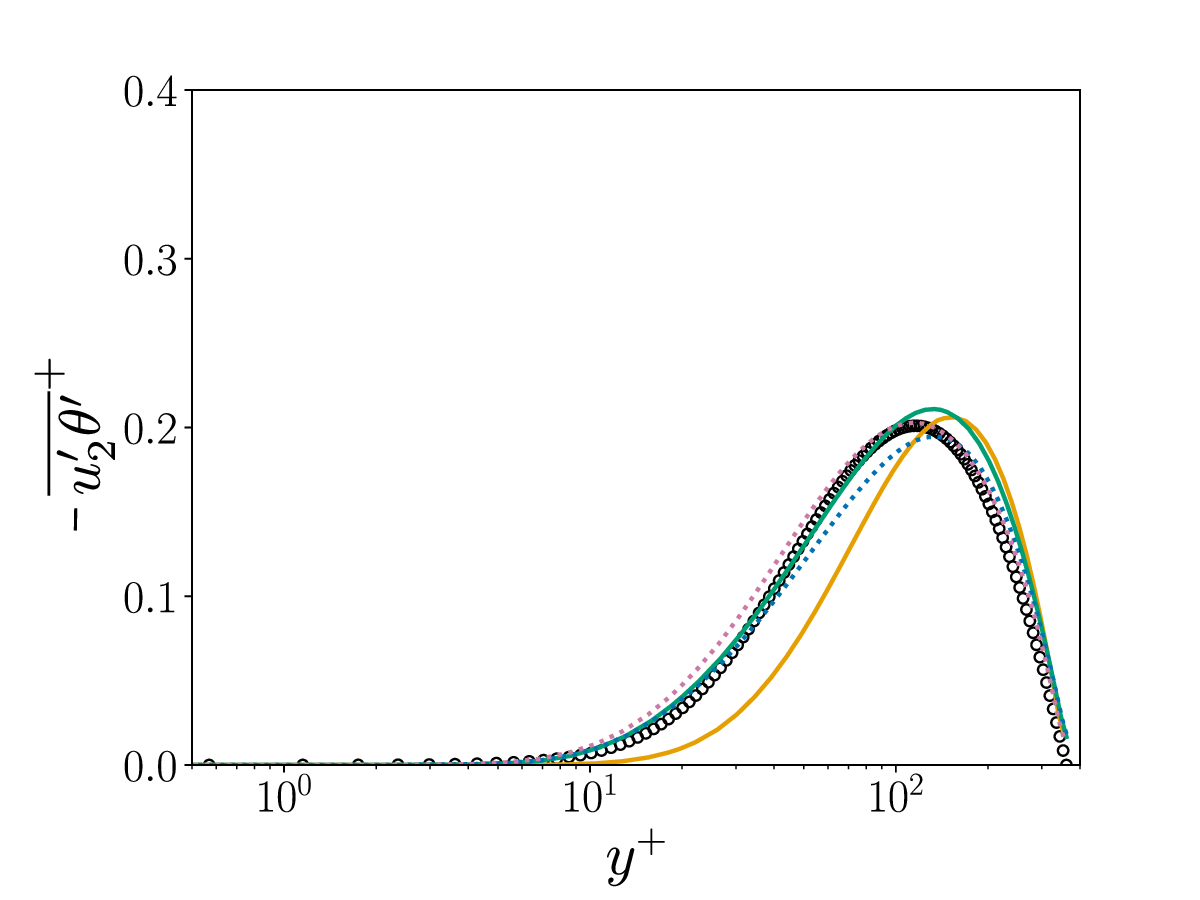}
        \caption{}
        \label{subfig:PF_ut}
    \end{subfigure}
    \hfill
    \begin{subfigure}{0.48\textwidth}
        \centering
        \includegraphics[width=\linewidth]{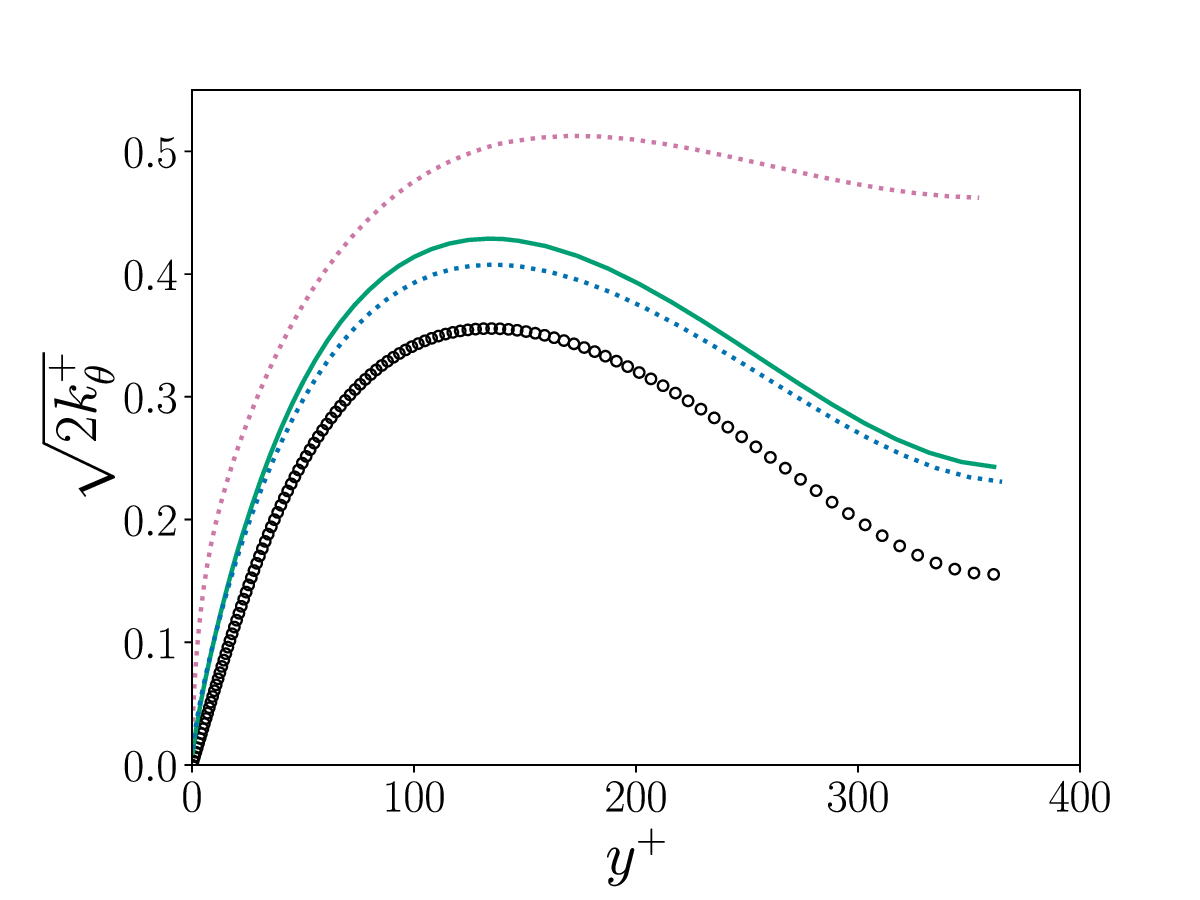}
        \caption{}
        \label{subfig:PF_kTheta}
    \end{subfigure}\vspace{0.5cm}
    
              % Legend included in first figure
\renewcommand{\arraystretch}{1.1}\tiny
\begin{tabular}{|>{\centering\arraybackslash}p{0.1cm}>{\centering\arraybackslash}p{2cm}|
                >{\centering\arraybackslash}p{0.1cm}>{\centering\arraybackslash}p{2cm}|}
    \hline
    \textcolor[HTML]{E69F00}{\rule[0.5ex]{0.3cm}{1.0pt}} & KWSST-KAYS &
    \textcolor[HTML]{009E73}{\rule[0.5ex]{0.3cm}{1.0pt}} & AKN-MM \\
    \hline
    \textcolor[HTML]{D55E00}{\rule[0.5ex]{0.3cm}{1.0pt}} & ShamsKE-AHFM &
    \simFalseSmall & LES \\
    \hline
    \tikz[baseline=-0.5ex]{\draw[colKLW, dotted, line width=1.2pt] (0,0) -- (0.3,0);} & KLW-DAVIA &
    \tikz[baseline=-0.5ex]{\draw[colEBRSM, dotted, line width=1.2pt] (0,0) -- (0.3,0);} & EBRSM-AHFM \\
    \hline
\end{tabular}
    \caption{(a) streamwise velocity $u_1^+ = u_1/u_\tau$ (b) temperature $\theta^+ = \theta/T_\tau$, (c) wall-normal turbulent
heat flux $-\overline{u_2^\prime \theta^\prime}^+ = -\overline{u_2^\prime \theta^\prime}/(u_\tau T_\tau)$ (d) temperature standard deviation \(\sqrt{2k_\theta}^+ = \sqrt{2k_\theta}/T_\tau\). LES data from \citet{StraubDataBase}.}
    \label{fig:PF_u_T_vT_kTheta}
\end{figure}

\Cref{fig:PF_u_T_vT_kTheta} presents the high-resolution LES data from
\citet{StraubDataBase} alongside the RANS simulation results for
selected variables in wall units. The results show qualitative
similarities to those observed in channel flow at
\(Re_\tau = 395\), thus only the key
differences between the models will be discussed rather than a detailed
profile-by-profile comparison.

The reproducibility of the implemented models is first assessed by
comparison with results available in the literature. For the AKN-MM and KLW-DAVIA combinations, \citet{Manservisi2014} and \citet{Manservisi2016} report
profiles of \(\theta^+\) at \(Re_\tau = 395\) and \(Pr = 0.025\),
although only in logarithmic representation, which limits a direct
quantitative comparison. However, \citet{Manservisi2014} provide
temperature standard deviation profiles with linear scaling, enabling a
more meaningful assessment.

For the present pipe flow configuration at \(Re_b = 11700\), the MM and
DAVIA models yield an estimated \(Re_\tau \approx 370\). At this value,
the AKN-MM combination slightly overestimates \(\sqrt{2k_\theta^+}\) compared to
the results reported by \citet{Manservisi2014} at
\(Re_\tau = 395\). A similar trend was already observed in the channel
flow case, indicating a systematic behaviour of the model. Despite these
differences, the overall profile is well captured. For the DAVIA model,
no reference data for \(\sqrt{2k_\theta^+}\) in cylindrical geometries
are available in \citet{Manservisi2016}, preventing a direct comparison.

Following this verification, the models are compared against the reference data of \cite{StraubDataBase}.
As shown in \Cref{fig:PF_u_T_vT_kTheta}, all models accurately reproduce
the mean velocity profile. The turbulent heat flux is also well captured
by all models, with the exception of KWSST-KAYS, which slightly underestimates it outside the
core region. As discussed in \Cref{subsec:channel}, at $Pr=0.025$ this has only a minor
effect on the mean temperature due to the dominance of molecular conduction.
Consequently, all models predict the mean temperature profile with minor
deviations, primarily in the core flow region.

As discussed in \Cref{subsec:channel}, at this Prandtl number bulk metrics such as the
Nusselt number are only weakly diagnostic of the accuracy of the modeled
turbulent heat flux; for this reason, no Nusselt-number comparison is
reported for the pipe flow, and the profile-resolved comparisons above are
retained as the primary basis for assessing model accuracy.

Regarding temperature fluctuations, the AKN-MM and KLW-DAVIA combinations reproduce
the overall behaviour of the LES data, although a slight overestimation
of the temperature standard deviation is observed, consistent with the
channel flow results. The EBRSM-AHFM model exhibits a more pronounced
overestimation, in line with the trends already identified in the
channel configuration.

In terms of numerical robustness, no significant differences are
observed compared to the channel flow simulations. The convergence
behaviour and sensitivity to initial conditions remain consistent across
both geometries.

In both channel and pipe flows, only one component of the Reynolds
stress tensor and turbulent heat flux directly affects the first-order
moments. As a result, these configurations provide limited insight into
the full behavior of the models. To enable a more comprehensive
assessment, a flow case with multi-directional gradients and coupled
interactions is required. For this purpose, the models are next
evaluated in a two-dimensional backward-facing step configuration in the following section. 

\subsection{Backward Facing Step (BFS)}
\label{subsec:bfs}
The backward-facing step (BFS) configuration is considered to assess
the performance of the turbulence models in a more complex flow
geometry compared to channel and pipe flows. In addition, the
simulations are performed at a lower Prandtl number (\(Pr = 0.0088\)),
selected to match the available DNS reference data of \citet{Niemann2016}.

Available reference results for this configuration are limited.   \citet{Schumm2015} examined the AKN-MM combination and KWSST-KAYS, while \citet{Shams2018Number3} considered the ShamsKE-AHFM combination. No reference data are available for the DAVIA and EBRSM-AHFM combinations.  Despite the discrepancies observed in channel flow, the ShamsKE-AHFM
combination is included in order to verify whether the lack of
reproducibility is systematic, consistently with the rationale discussed in \Cref{subsec:channel}.

\begin{figure}
\centering
\def\svgwidth{0.75\columnwidth}
% ===== begin BFS_sketch.pdf_tex =====
%% Creator: Inkscape 1.4.3 (1:1.4.3+202512261034+0d15f75042), www.inkscape.org
%% Accompanies image file 'BFS_sketch.pdf' (pdf, eps, ps)
\begingroup%
  \makeatletter%
  \providecommand\color[2][]{%
    \errmessage{(Inkscape) Color is used for the text in Inkscape, but the package 'color.sty' is not loaded}%
    \renewcommand\color[2][]{}%
  }%
  \providecommand\transparent[1]{%
    \errmessage{(Inkscape) Transparency is used (non-zero) for the text in Inkscape, but the package 'transparent.sty' is not loaded}%
    \renewcommand\transparent[1]{}%
  }%
  \providecommand\rotatebox[2]{#2}%
  \newcommand*\fsize{\dimexpr\f@size pt\relax}%
  \newcommand*\lineheight[1]{\fontsize{\fsize}{#1\fsize}\selectfont}%
  \ifx\svgwidth\undefined%
    \setlength{\unitlength}{679.3507157bp}%
    \ifx\svgscale\undefined%
      \relax%
    \else%
      \setlength{\unitlength}{\unitlength * \real{\svgscale}}%
    \fi%
  \else%
    \setlength{\unitlength}{\svgwidth}%
  \fi%
  \global\let\svgwidth\undefined%
  \global\let\svgscale\undefined%
  \makeatother%
  \begin{picture}(1,0.57278704)%
    \lineheight{1}%
    \setlength\tabcolsep{0pt}%
    \put(0,0){\includegraphics[width=\unitlength]{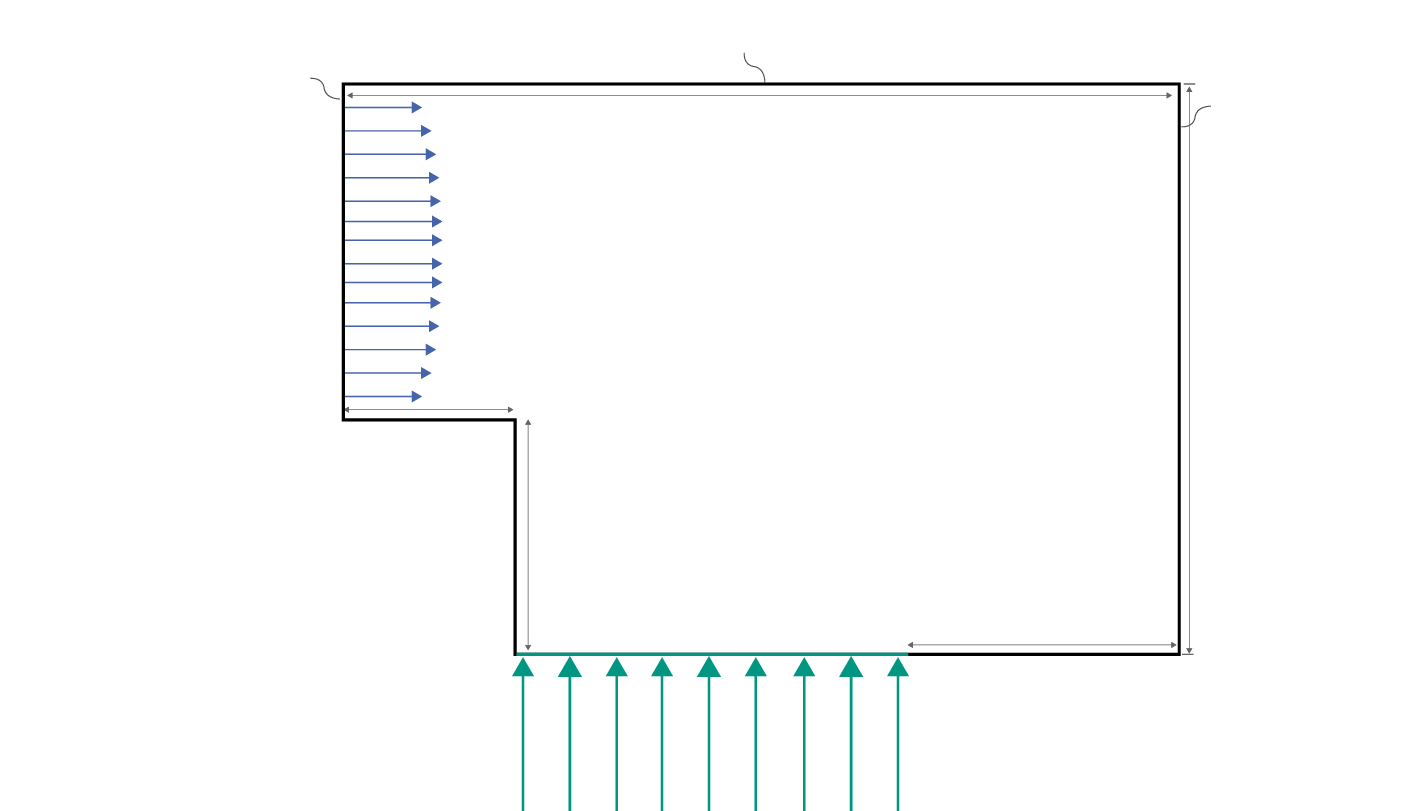}}%
    \put(0.42815024,0.54318167){\color[rgb]{0,0,0}\makebox(0,0)[lt]{\lineheight{1.25}\smash{\begin{tabular}[t]{l}$\text{adiabatic wall}$\end{tabular}}}}%
    \put(0,0){\includegraphics[width=\unitlength]{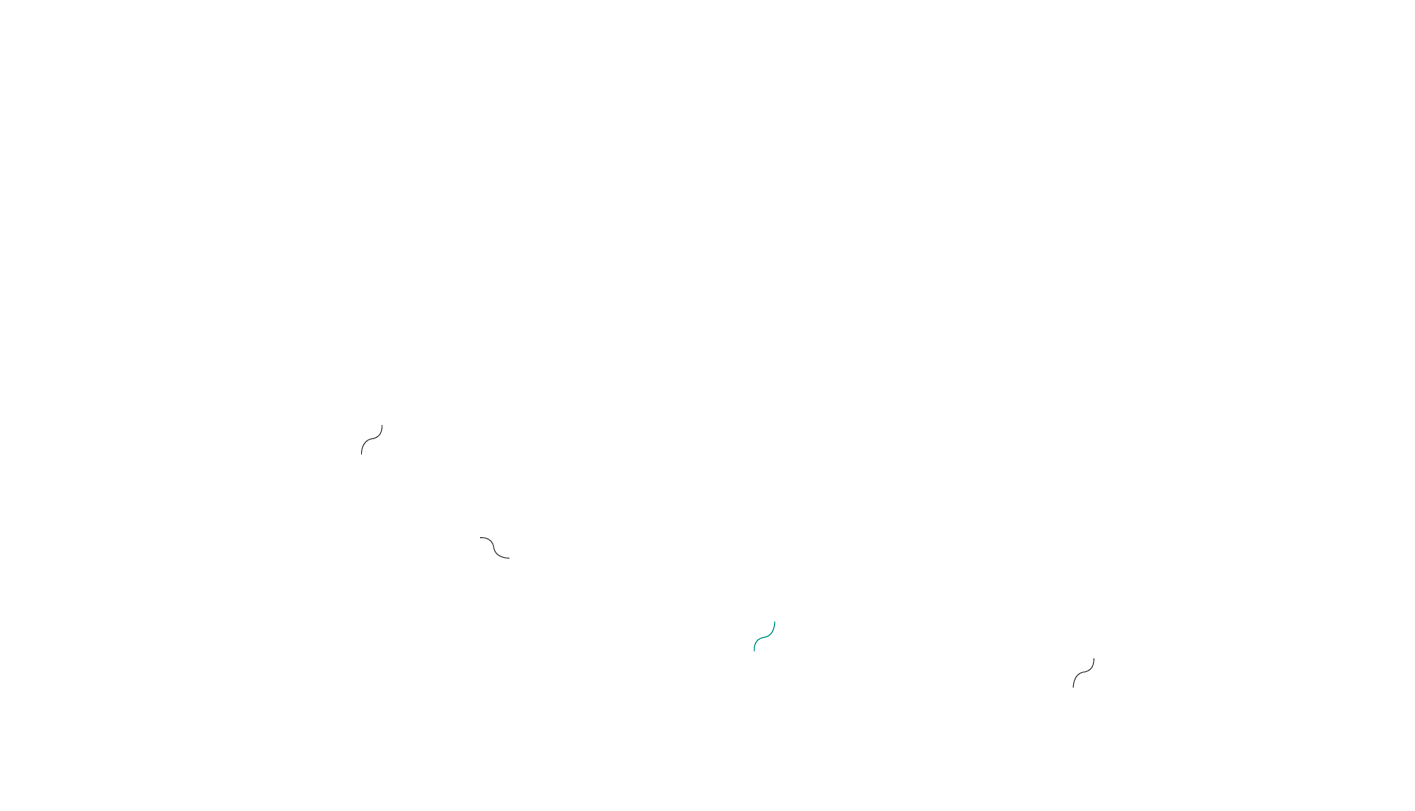}}%
    \put(0.38591827,0.13973395){\color[rgb]{0,0,0}\makebox(0,0)[lt]{\lineheight{1.25}\smash{\begin{tabular}[t]{l}$\textcolor[RGB]{0, 150, 130}{\text{heated wall}}$\end{tabular}}}}%
    \put(0,0){\includegraphics[width=\unitlength]{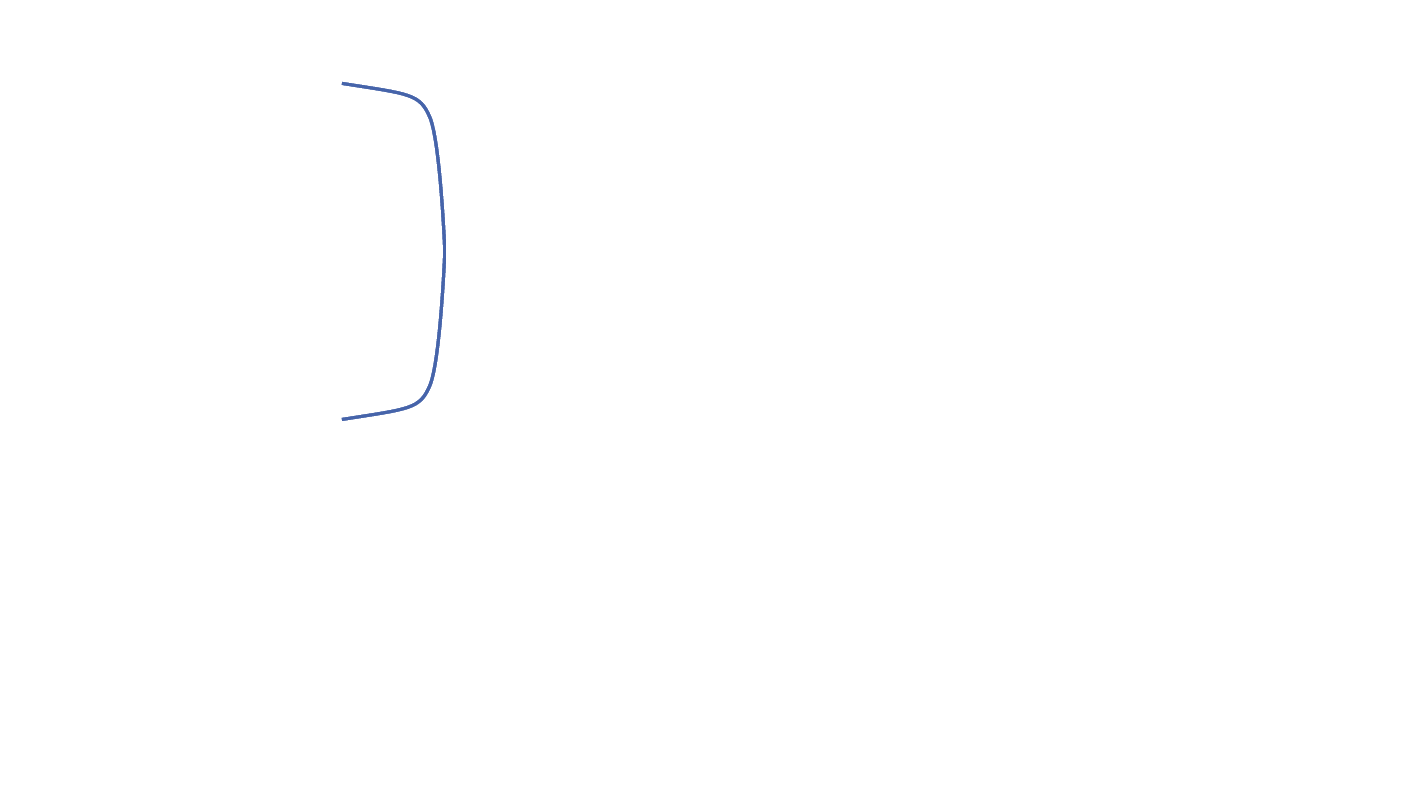}}%
    \put(0.21129394,0.40453426){\color[rgb]{0,0,0}\rotatebox{90}{\makebox(0,0)[lt]{\lineheight{1.25}\smash{\begin{tabular}[t]{l}$\text{inlet}$\end{tabular}}}}}%
    \put(0.89376003,0.4024579){\color[rgb]{0,0,0}\rotatebox{90}{\makebox(0,0)[lt]{\lineheight{1.25}\smash{\begin{tabular}[t]{l}$\text{outlet}$\end{tabular}}}}}%
    \put(-0.00141555,0.20861209){\color[rgb]{0,0,0}\makebox(0,0)[lt]{\lineheight{1.25}\smash{\begin{tabular}[t]{l}$\text{adiabatic wall}$\end{tabular}}}}%
    \put(0.67666035,0.04303979){\color[rgb]{0,0,0}\makebox(0,0)[lt]{\lineheight{1.25}\smash{\begin{tabular}[t]{l}$\text{adiabatic wall}$\end{tabular}}}}%
    \put(0,0){\includegraphics[width=\unitlength]{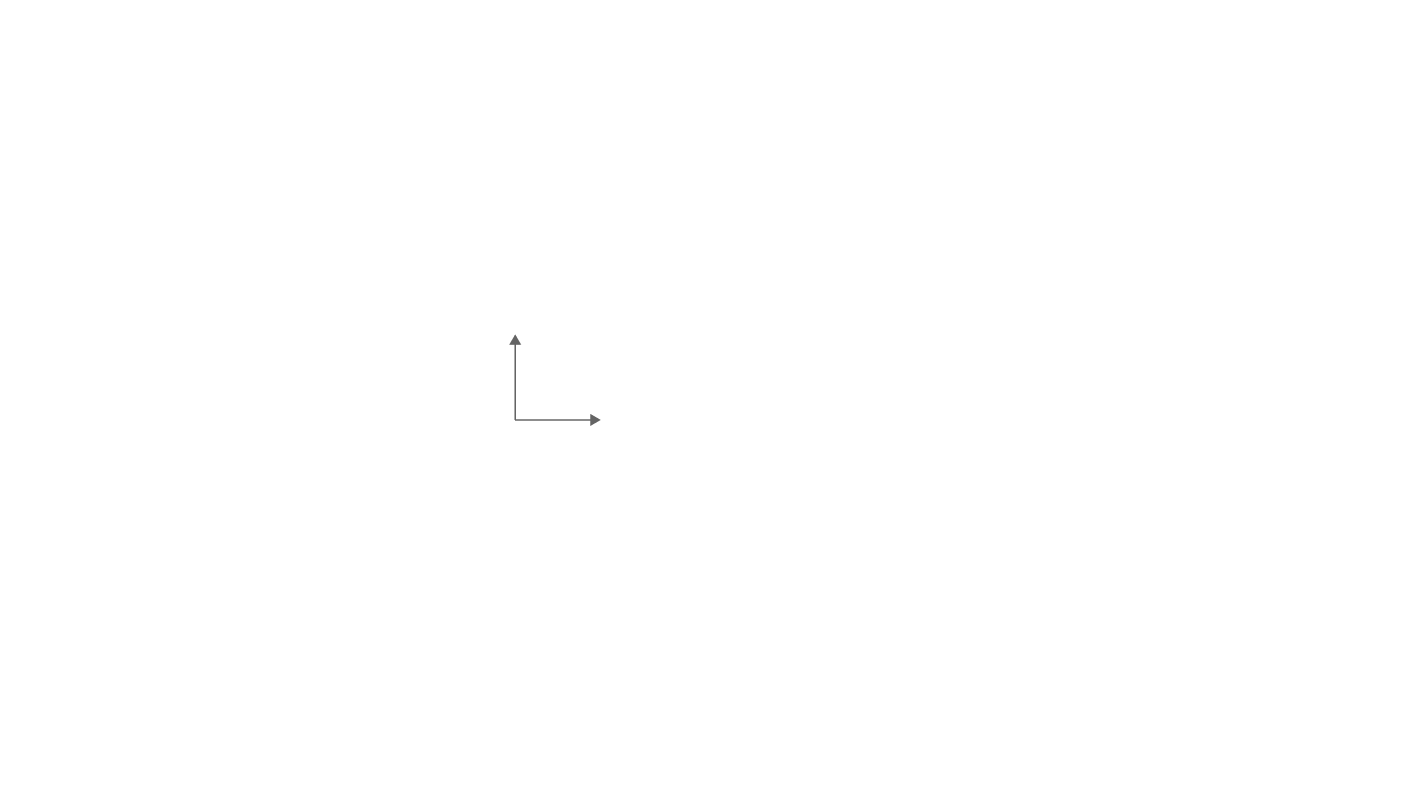}}%
    \put(0.42197326,0.28378257){\color[rgb]{0.4,0.4,0.4}\makebox(0,0)[lt]{\lineheight{1.25}\smash{\begin{tabular}[t]{l}$\textcolor[RGB]{100,100,100}{e_x}$\end{tabular}}}}%
    \put(0.36792081,0.3563867){\color[rgb]{0.4,0.4,0.4}\makebox(0,0)[lt]{\lineheight{1.25}\smash{\begin{tabular}[t]{l}$\textcolor[RGB]{100,100,100}{e_y}$\end{tabular}}}}%
    \put(0.3898654,0.24474311){\color[rgb]{0.4,0.4,0.4}\makebox(0,0)[lt]{\lineheight{1.25}\smash{\begin{tabular}[t]{l}$\textcolor[RGB]{100,100,100}{e_z}$\end{tabular}}}}%
    \put(0,0){\includegraphics[width=\unitlength]{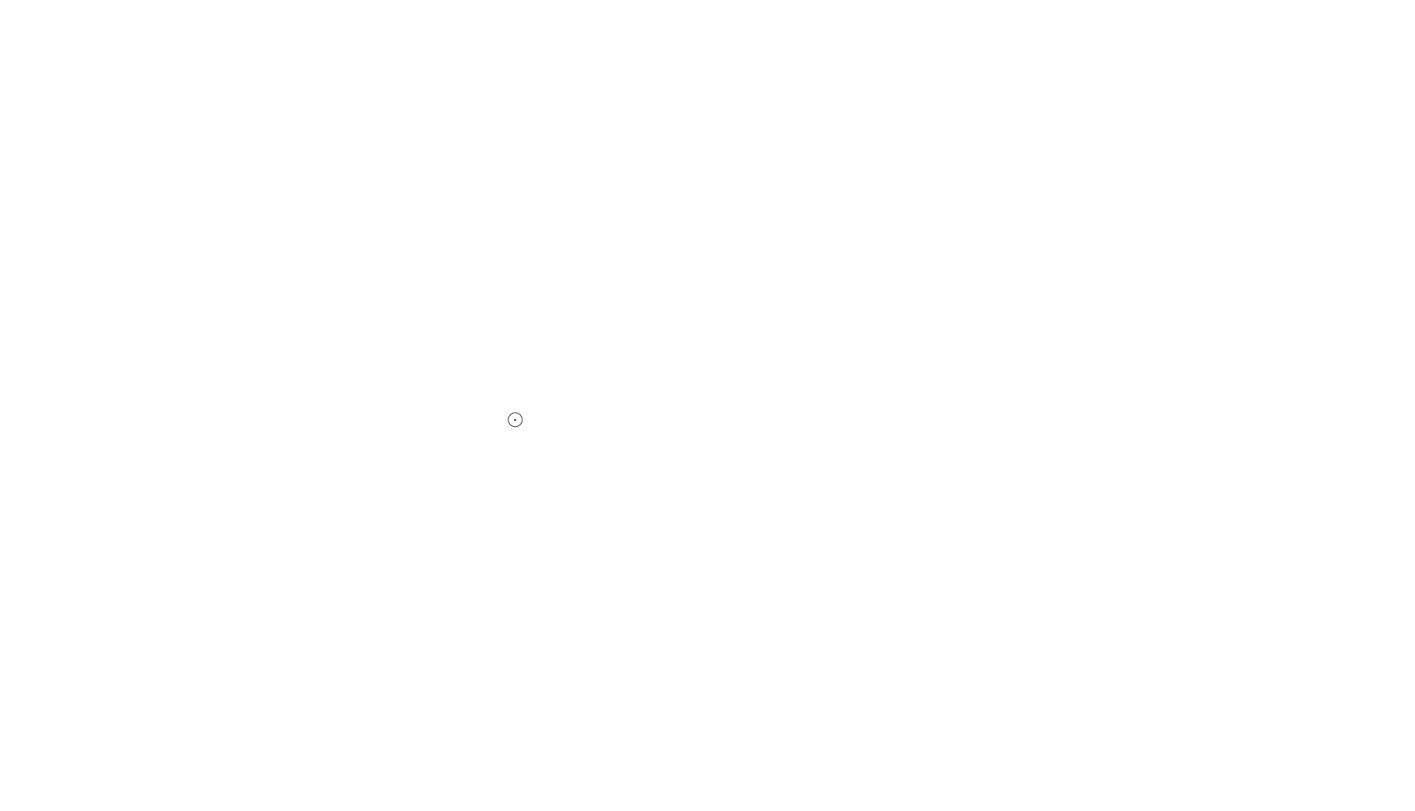}}%
    \put(0.30392935,0.0471882){\color[rgb]{0,0,0}\makebox(0,0)[lt]{\lineheight{1.25}\smash{\begin{tabular}[t]{l}$\textcolor[RGB]{0, 150, 130}{q_w}$\end{tabular}}}}%
    \put(0.31695194,0.46421926){\color[rgb]{0,0,0}\makebox(0,0)[lt]{\lineheight{1.25}\smash{\begin{tabular}[t]{l}$\textcolor[RGB]{70, 100, 170}{u_{1,\text{ch}}}$\end{tabular}}}}%
    \put(0.48759167,0.47207462){\color[rgb]{0.4,0.4,0.4}\makebox(0,0)[lt]{\lineheight{1.25}\smash{\begin{tabular}[t]{l}$\textcolor[RGB]{100,100,100}{l_{top}}$\end{tabular}}}}%
    \put(0.84995515,0.30775674){\color[rgb]{0.4,0.4,0.4}\makebox(0,0)[lt]{\lineheight{1.25}\smash{\begin{tabular}[t]{l}$\textcolor[RGB]{100,100,100}{L_{step}}$\end{tabular}}}}%
    \put(0.38063525,0.19376388){\color[rgb]{0.4,0.4,0.4}\makebox(0,0)[lt]{\lineheight{1.25}\smash{\begin{tabular}[t]{l}$\textcolor[RGB]{100,100,100}{l_{step}}$\end{tabular}}}}%
    \put(0.71106853,0.12931894){\color[rgb]{0.4,0.4,0.4}\makebox(0,0)[lt]{\lineheight{1.25}\smash{\begin{tabular}[t]{l}$\textcolor[RGB]{100,100,100}{l_{out}}$\end{tabular}}}}%
    \put(0.31042518,0.29876805){\color[rgb]{0.4,0.4,0.4}\makebox(0,0)[lt]{\lineheight{1.25}\smash{\begin{tabular}[t]{l}$\textcolor[RGB]{100,100,100}{l_{in}}$\end{tabular}}}}%
  \end{picture}%
\endgroup%
% ===== end BFS_sketch.pdf_tex =====
% \includegraphics[width=\linewidth]{BFS-sketch.eps}
\caption{Schematic (not to scale) of the backward-facing step configuration.
\(u_{1,\text{ch}}\) represents the prescribed inlet velocity
profile obtained from a fully developed turbulent channel flow.}
\label{fig:BFS_sketch}
\end{figure}
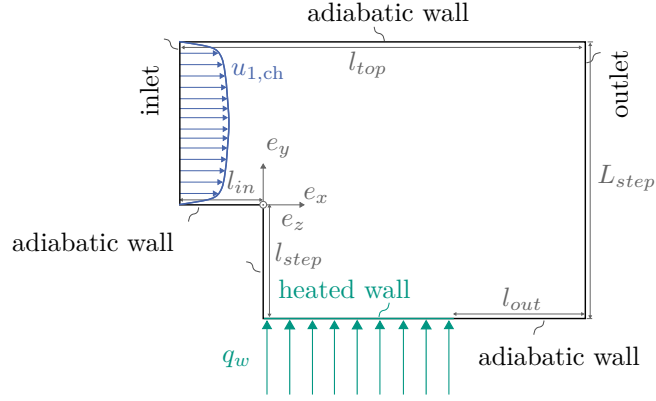

The BFS geometry is illustrated in \Cref{fig:BFS_sketch}. The streamwise dimensions of the domain are defined relative to the
step height \(l_{\mathrm{step}}\), with
\(l_{\mathrm{in}}/l_{\mathrm{step}} = 2\),
\(l_{\mathrm{out}}/l_{\mathrm{step}} = 10\), and
\(l_{\mathrm{top}}/l_{\mathrm{step}} = 22\).
The expansion ratio is \(ER = L_{\mathrm{step}} /
(L_{\mathrm{step}} - l_{\mathrm{step}}) = 1.5\). The setup follows that of \citet{Shams2018Number3}, featuring an adiabatic section absent in \citet{Niemann2016}. The computational mesh consists of 128,800 cells, refined to ensure $y^+ < 1$ across all cases.

Following \citet{Niemann2016}, the inlet bulk Reynolds number is set to $Re_b = 2\,l_\mathrm{step}u_b/\nu = 9610$. Velocity and momentum turbulence variables are initialized using
profiles obtained from fully developed channel flow simulations,
as exemplified by the inlet profile $u_{1,\mathrm{ch}}$ shown in
\Cref{fig:BFS_sketch}. The pressure at the inlet has a zero-gradient condition, while the temperature is fixed at $T_\mathrm{ref}$. According to \citet{Schumm2015}, small initial values are prescribed for $k_\theta$ and $\epsilon_\theta$, with $\Omega_\theta$ computed accordingly.

At the outlet, a Dirichlet condition of zero pressure is applied, while other variables follow zero-gradient conditions. A portion of the downstream wall is heated, while other sections remain adiabatic. A constant and uniform heat flux $q_w$ is applied at the heated section. 
% is $\bar{\dot{q}}_w = 41$ kW/m$^2$ based on \citet{Schumm2015}.

For the EBRSM-AHFM model, \(C_{t1} = 0.0052\) is used following
\citet{Shams2018Number3}, instead of the value \(C_{t1} = 0.25\)
obtained from \Cref{eqn:AHFM2014} or \Cref{eqn:AHFM2019}. This choice
stems from the formulation adopted by \citet{Shams2018Number3},
where, for \(Re\,Pr < 180\), the value \(Re\,Pr = 180\) is used in
\Cref{eqn:AHFM2014}, leading to a reduced value of \(C_{t1}\).

Steady-state simulations are performed with constant fluid properties.  Central-difference schemes are applied to the
diffusive terms, while linear upwind schemes are used for the advective terms. The energy equation is solved using a \textit{frozen-flow} approach after convergence of the velocity field. For the thermal turbulence models MM, DAVIA, and AHFM, the initial temperature field is obtained from
simulations with a constant turbulent Prandtl number equal to 1.
Convergence is assumed when the residuals of the thermal variables drop below \(10^{-6}\).

For the EBRSM model, no converged turbulent solution could be obtained for any of the tested
numerical settings, despite using different orders of magnitude of \(\nu\),
various initial conditions, numerical schemes, and equation
under-relaxation factors. Consequently, no results are presented for
the EBRSM-AHFM combination.

It should be noted that converged EBRSM solutions for BFS
and separated flows have been reported in the literature with other codes,
notably with the reference implementation in Code\_Saturne \cite{Manceau2015,Shams2019Number3}. This
supports the interpretation that the convergence issues observed here are
related to the numerical treatment of the present OpenFOAM implementation
rather than to the model formulation itself. A comparison with the
alternative OpenFOAM implementation recently released by \citet{marocco_2024} is envisaged to isolate the origin of these issues.

A converged turbulent solution for the momentum-related variables could
be obtained using the KLW model. However, all simulations involving the
thermal turbulence model KLW-DAVIA diverged. This behaviour may be attributed to the
cross-diffusion terms appearing in the transport equations of the
thermal turbulence variables, as described in \Cref{subsubsec:kThetaOmegaTheta}. For sufficiently large gradients, these terms can become dominant compared to the other contributions in the transport
equations, potentially leading to divergence.
Since no converged solution could be achieved for the thermal turbulence
model, only the results of the KLW momentum model are presented.

\begin{figure}
  \centering
    \begin{subfigure}[b]{0.9\textwidth}
        \centering     
    \includegraphics[width=1.0\linewidth]{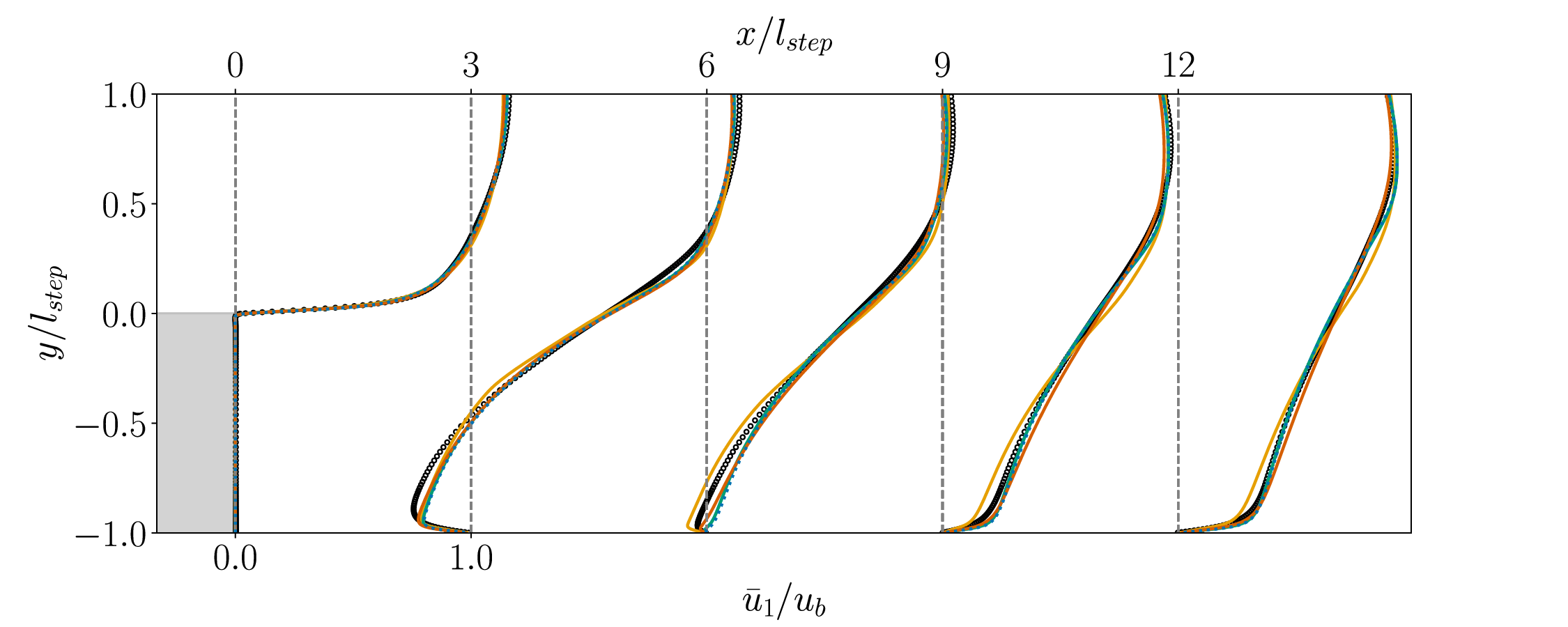}
    \caption{}
      \label{subfig:BFS_u1}  
      
    \end{subfigure}
    \begin{subfigure}[b]{0.9\textwidth}
        \centering
      \includegraphics[width=1.0\linewidth]{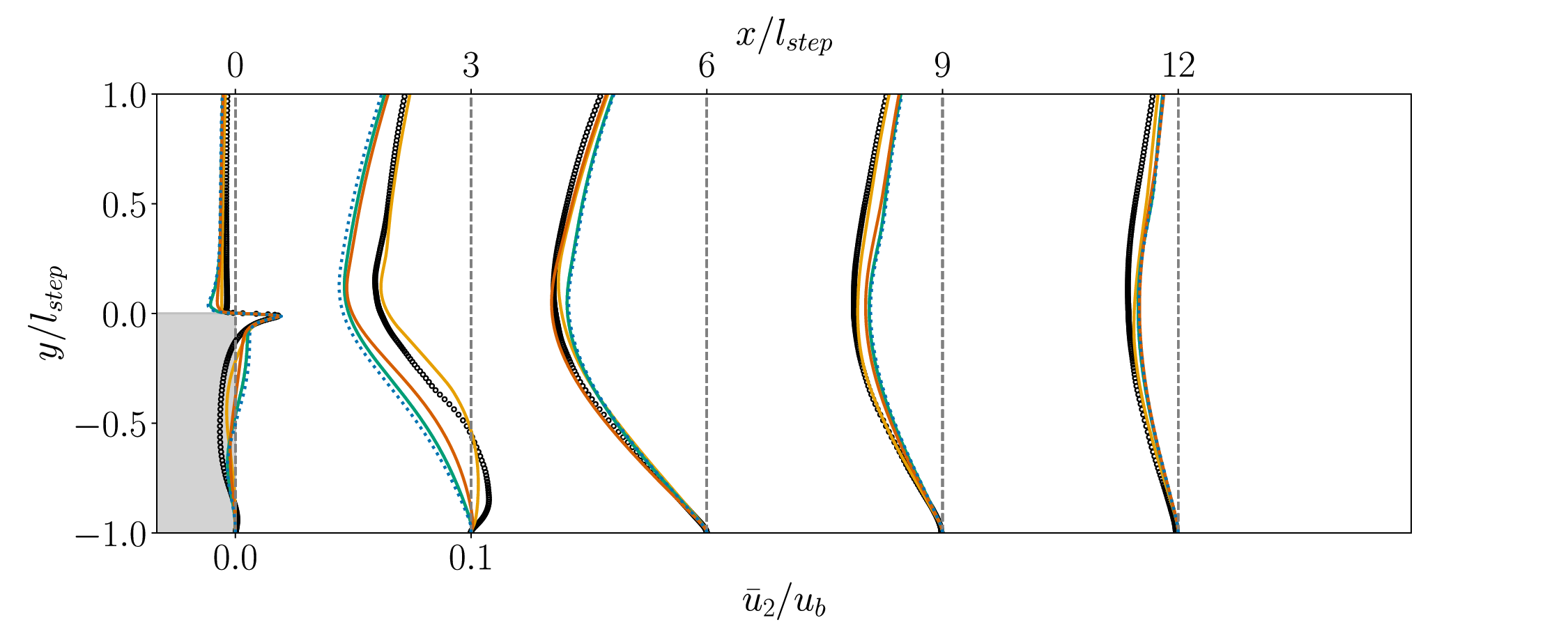}
      \caption{}
      \label{subfig:BFS_u2}
  \end{subfigure}

    \begin{subfigure}[b]{0.9\textwidth}
        \centering
        \includegraphics[width=0.5\linewidth]{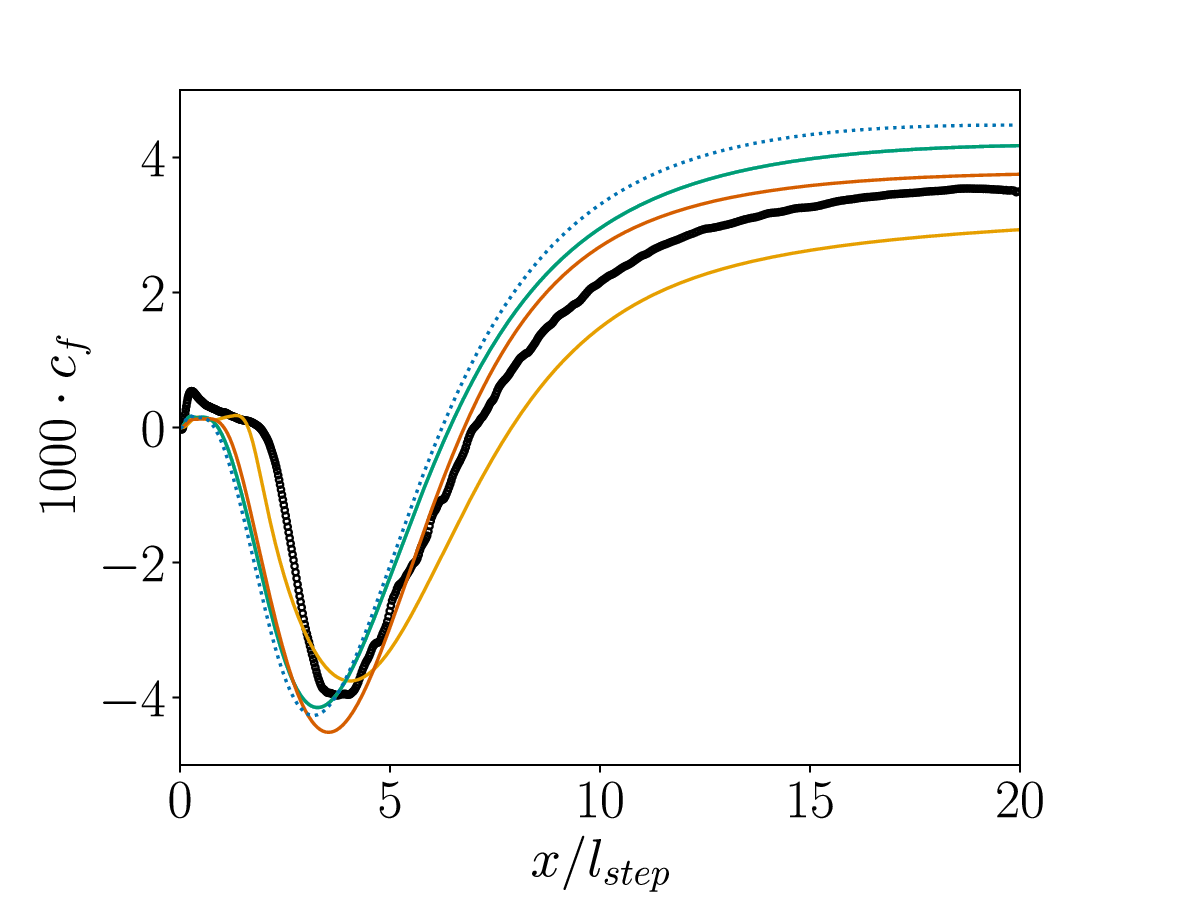}
        \caption{}
      \label{subfig:BFS_cf}
  \end{subfigure}\vspace{0.5cm}
  % Legend
        \renewcommand{\arraystretch}{1.1}\tiny
        \begin{tabular}{|>{\centering\arraybackslash}p{0.1cm}>{\centering\arraybackslash}p{1.0cm}|>{\centering\arraybackslash}p{0.1cm}>{\centering\arraybackslash}p{1.0cm}|>{\centering\arraybackslash}p{0.06cm}>{\centering\arraybackslash}p{0.7cm}|}
            \hline
            \textcolor[HTML]{E69F00}{\rule[0.5ex]{0.3cm}{1.0pt}} & KWSST & \textcolor[HTML]{D55E00}{\rule[0.5ex]{0.3cm}{1.0pt}} & ShamsKE  &\simFalseSmall & DNS\\
            \hline
            \textcolor[HTML]{009E73}{ \rule[0.5ex]{0.3cm}{1.0pt}} & AKN  & \textcolor[HTML]{0072B2}{\rule[0.5ex]{0.05cm}{1.0pt}}\textcolor[HTML]{FFFFFF}{\rule[0.5ex]{0.05cm}{1.0pt}}\textcolor[HTML]{0072B2}{\rule[0.5ex]{0.05cm}{1.0pt}}\textcolor[HTML]{FFFFFF}{\rule[0.5ex]{0.05cm}{1.0pt}}\textcolor[HTML]{0072B2}{\rule[0.5ex]{0.05cm}{1.0pt}} & KLW & & \\
            \hline
        \end{tabular}
  \caption{(a) streamwise velocity,
(b) wall-normal velocity,
(c) skin friction coefficient \(c_f\) along the heated wall section.
DNS data from
\cite{Niemann2016}.}
  \label{fig:BFS_momentum}
\end{figure}

The results of the turbulence models are presented in terms of the mean
velocity in the streamwise and wall-normal directions in
\Cref{subfig:BFS_u1,subfig:BFS_u2} at different downstream locations, up to a
distance of \(2/3\,L_{\mathrm{step}}\) from the lower wall. The skin
friction coefficient \(c_f\) along the heated wall is shown in
\Cref{subfig:BFS_cf}. Following \citet{Schumm2015}, it is defined as \( c_f = \tau_w/(0.5\rho u^2_{b,\text{in}})\), being $\tau_w$ the wall-shear stress and $u_{b,\text{in}}$ the inlet bulk velocity.

The solutions for the momentum-related quantities obtained with the KWSST and AKN models are consistent with the results reported by \citet{Schumm2015}. Similarly, the ShamsKE model reproduces the results of \citet{Shams2018Number3}. This provides further evidence that the implemented variant is consistent with
that used by \citet{Shams2014}, suggesting that the discrepancies
observed for the ShamsKE-AHFM combination in channel flow are likely related to the thermal turbulence model.

Since the KLW model is derived from the AKN model, their results show very similar trends, as illustrated in \Cref{subfig:BFS_u1} and \Cref{subfig:BFS_u2}. Owing to this similarity, the discussion focuses primarily on the AKN model. Given the good agreement with the results reported by the original authors, only selected aspects are discussed in the following.

As shown in \Cref{subfig:BFS_u1,subfig:BFS_u2}, the KWSST model
provides the best agreement with the DNS mean velocity upstream of the reattachment point, located at \(x/l_{\mathrm{step}} = 7.01\) according to \citet{Niemann2016}. The skin friction coefficient, shown in \Cref{subfig:BFS_cf}, is also well captured upstream of the reattachment point. Downstream thereof,  \(c_f\)  increases gradually, reflecting the development of the boundary layer, in agreement with \citet{Schumm2015}. Overall, the KWSST model accurately predicts the recirculation region.

According to \citet{Shams2018Number3}, the corresponding ShamsKE
model exhibits the largest deviation from DNS in terms of \(c_f\) in the region close to the step. In addition, the size of the corner vortex is underestimated, a behaviour that is also observed for the AKN model.
Both the AKN and ShamsKE models overpredict the velocity in the
 wall-normal direction at \(x/l_{\mathrm{step}} = 3\).

Downstream of the reattachment point, the ShamsKE model provides the best agreement with the DNS skin friction coefficient among all considered models. It also predicts the reattachment location with the smallest deviation.

An accurate prediction of heat transfer requires a reliable representation of the flow field. All momentum turbulence models capture the main features of the mean velocity field at least qualitatively. However, the AKN, KLW, and ShamsKE models show
noticeable deviations in the prediction of the recirculation region, while the KWSST model exhibits larger discrepancies in the downstream velocity field beyond the reattachment point.

\begin{figure}[!htb]
  \centering
    \begin{subfigure}[b]{0.9\textwidth}
        \centering     
    \includegraphics[width=1.0\linewidth]{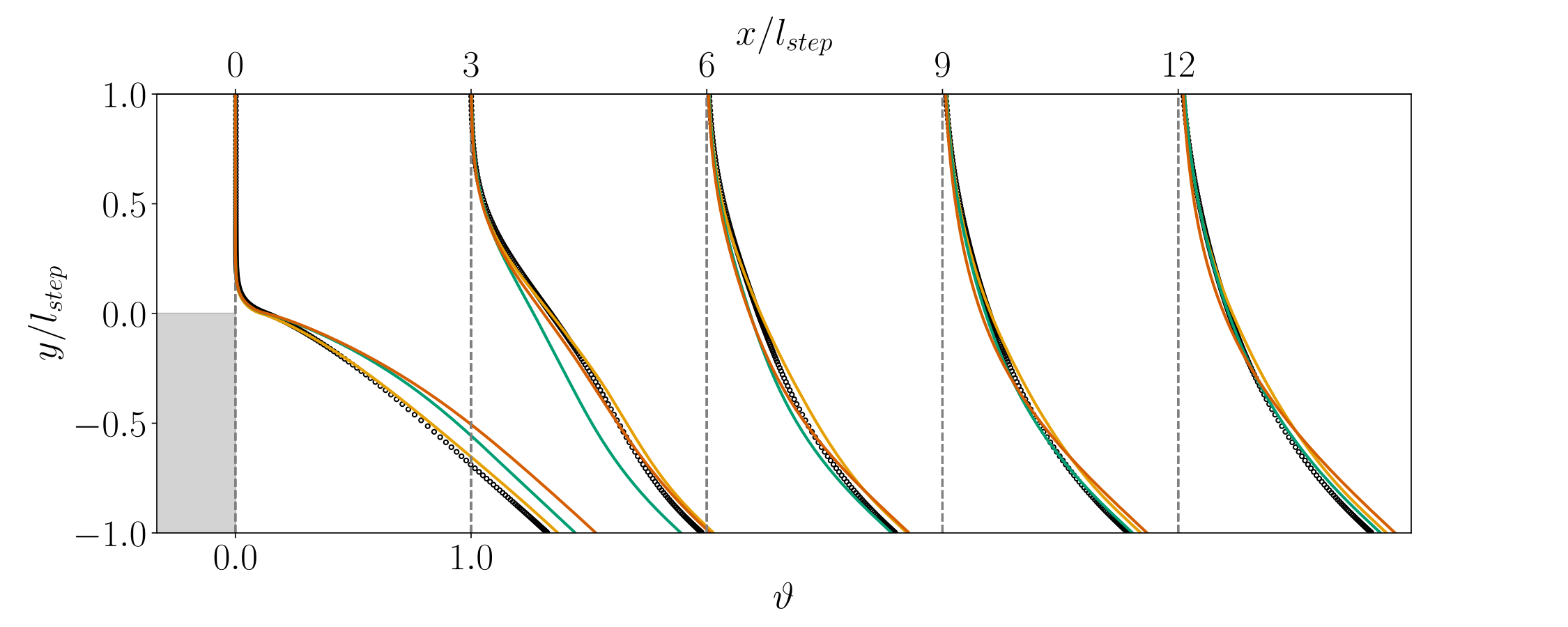}
    \caption{}
      \label{subfig:BFS_theta}  
      
    \end{subfigure}
    \begin{subfigure}[b]{0.9\textwidth}
        \centering
      \includegraphics[width=1.0\linewidth]{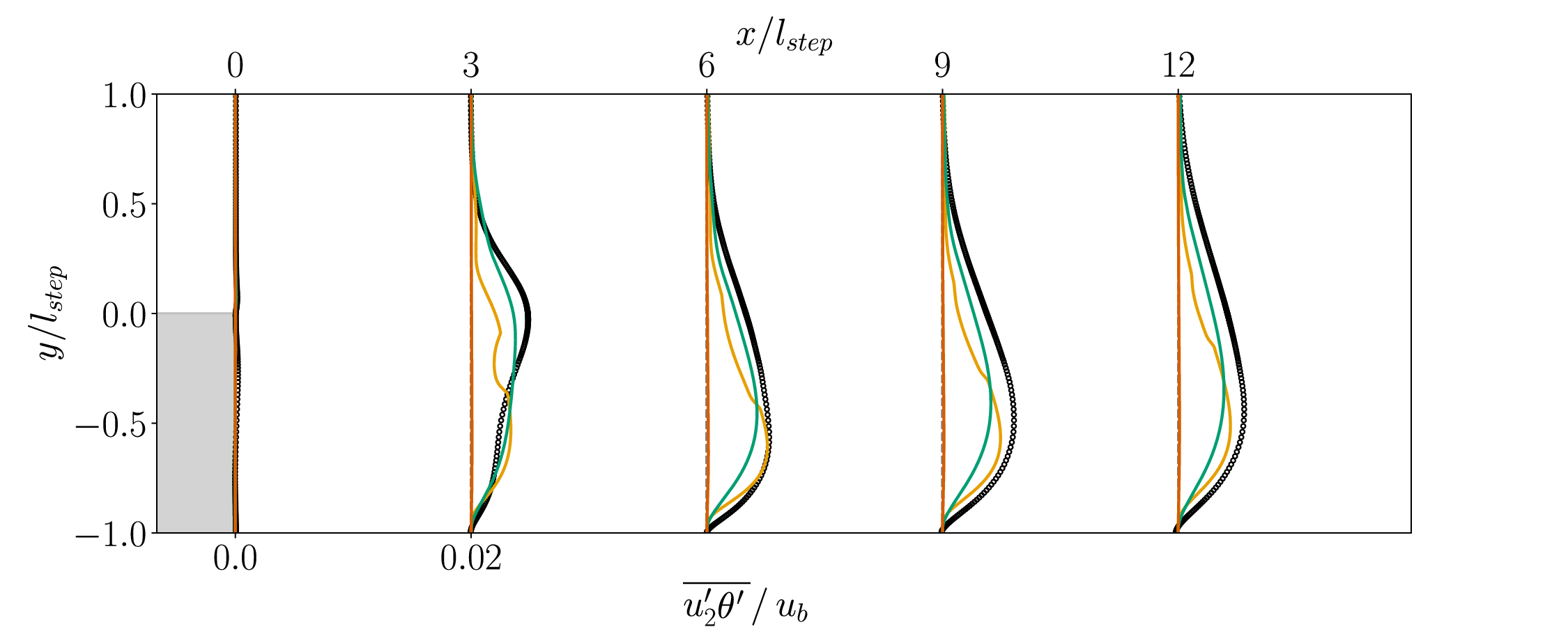}
      \caption{}
      \label{subfig:BFS_vT}
  \end{subfigure}

    \begin{subfigure}[b]{0.9\textwidth}
        \centering
        \includegraphics[width=0.5\linewidth]{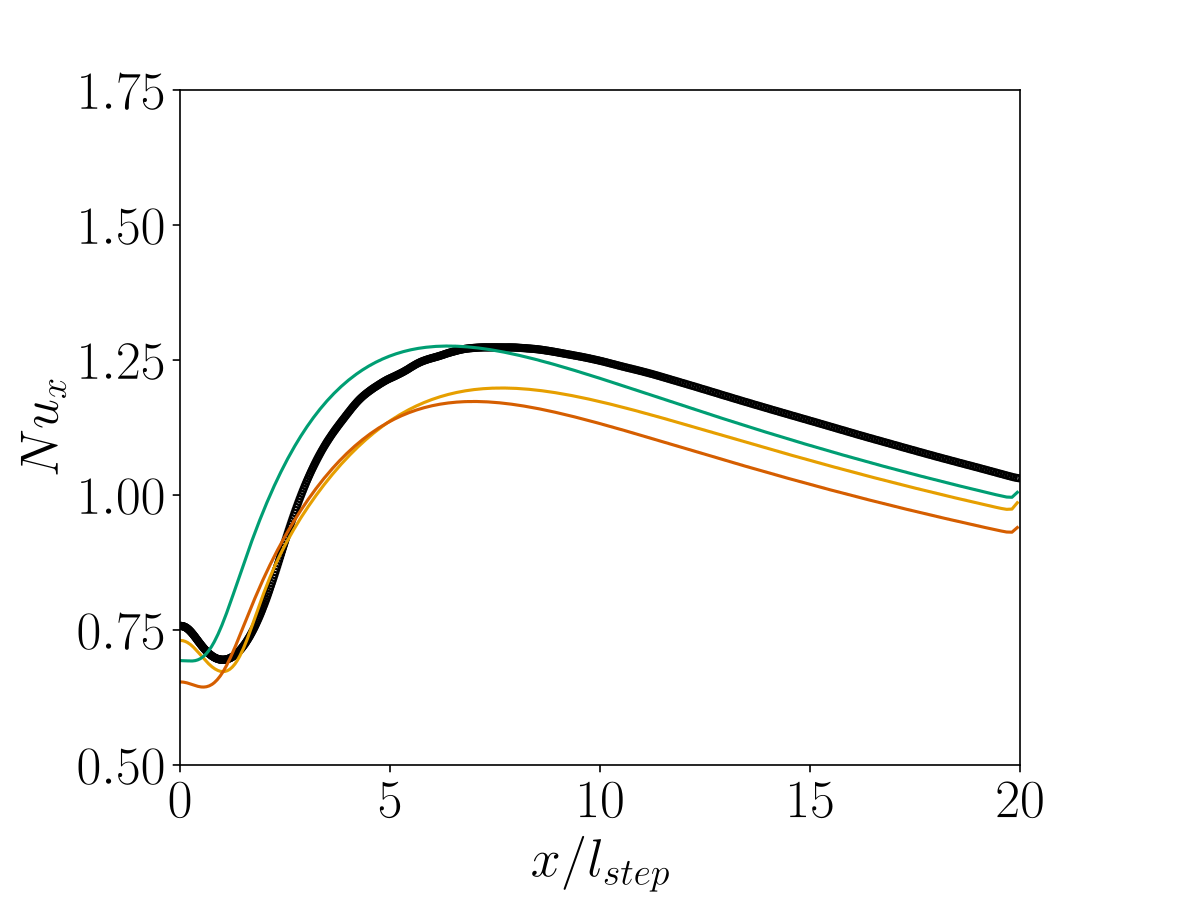}
        \caption{}
      \label{subfig:BFS_Nu}
  \end{subfigure}\vspace{0.5cm}
% Legend included in first figure
\renewcommand{\arraystretch}{1.1}\tiny
\begin{tabular}{|>{\centering\arraybackslash}p{0.1cm}>{\centering\arraybackslash}p{2cm}|>{\centering\arraybackslash}p{0.1cm}>{\centering\arraybackslash}p{2cm}|}
    \hline
    \textcolor[HTML]{E69F00}{\rule[0.5ex]{0.3cm}{1.0pt}} & KWSST-KAYS &
    \textcolor[HTML]{009E73}{\rule[0.5ex]{0.3cm}{1.0pt}} & AKN-MM \\
    \hline
    \textcolor[HTML]{D55E00}{\rule[0.5ex]{0.3cm}{1.0pt}} & ShamsKE-AHFM &
    \simFalseSmall & DNS \\
    \hline
\end{tabular}
  \caption{(a) excess temperature, (b) wall-normal turbulent heat flux, (c) local Nusselt number \(Nu_x\) along the heated wall section. DNS data from \cite{Niemann2016}.}
  \label{fig:BFS_thermal}
\end{figure}

% Thermal turbulence models
The results for selected thermal quantities are presented in
\Cref{fig:BFS_thermal}. Following \citet{Niemann2016}, the excess temperature \(\theta = (T - T_{\mathrm{ref}})/\Delta T\) is used, with \(\Delta T = q_w l_{\mathrm{step}} / \lambda\), being $\lambda$ the fluid's thermal conductivity.
The distribution of the excess temperature is shown in
\Cref{subfig:BFS_theta}, while profiles of the wall-normal turbulent heat flux at different downstream locations are reported in \Cref{subfig:BFS_vT}, up to a distance of
\(2/3\,L_{\mathrm{step}}\) from the lower wall.

The AKN-MM combination reproduces the results reported by
\citet{Schumm2015}, and reference is therefore made to their discussion
for a more detailed interpretation. Consistent with the channel flow
case, the ShamsKE-AHFM combination fails to reproduce the
temperature-related results reported by \citet{Shams2018Number3}.
Since the momentum part of this combination (ShamsKE) does
reproduce the reference velocity and skin-friction results discussed
above, this confirms that the lack of reproducibility originates
specifically from the thermal closure, and is systematic across flow
configurations rather than case-dependent. For this reason, the
corresponding temperature-related results are not discussed further in
the following.

As shown in \Cref{subfig:BFS_vT}, the wall-normal turbulent heat flux is
generally underestimated by the AKN-MM and KWSST-KAYS
combinations, particularly downstream of the reattachment point. This
leads to corresponding deviations in the predicted temperature field,
especially in the near-wall region.

While, in channel flow, discrepancies in the temperature field can be
largely attributed to inaccuracies in the wall-normal turbulent heat
flux, this simplification does not hold for the BFS case. Here, the
temperature field is influenced not only by \(\overline{u_2'\theta'}\), but also by the
accuracy of the predicted velocity field, due to the two-dimensional
nature of the flow and the presence of strong spatial gradients.

Furthermore, the dependence of the thermal turbulence models on the
underlying momentum model complicates the interpretation of the
results. In particular, since the MM model is combined with the AKN
momentum model, part of the observed differences can be attributed to the
interaction between the thermal and momentum closures rather than to
the thermal model alone.

For a clearer representation of the wall temperature, the local
Nusselt number \(Nu_x\) along the heated wall is shown in
\Cref{subfig:BFS_Nu}. Following \citet{Schumm2015}, it is defined as
\begin{equation}
    Nu_x = \frac{q_w \, l_{\mathrm{step}}}
    {\lambda \left(T_w - T_{\mathrm{ref}}\right)} = \frac{1}{\theta_w}
\end{equation}

Thus, \(Nu_x\) is inversely proportional to the excess temperature
at the wall. For the DNS data, the maximum wall temperature occurs
within the recirculation region, a behaviour that is also captured by
all models.

In the region close to the step, the KWSST-KAYS combination provides
the best agreement with the DNS \(Nu_x\) distribution. However,
downstream, it overpredicts the wall temperature. This can be attributed
to the overestimation of the recirculation length, which delays the
recovery of the boundary layer and reduces the wall temperature due to
enhanced mixing with colder fluid.
The AKN-MM combination overestimates \(Nu_x\) within the
recirculation region but provides the best agreement in the downstream
region compared to the other models.

\FloatBarrier
% ===== end resultsAndDiscussion.tex =====

\section{Summary and conclusions}
\label{sec:conclusions}
% ===== begin conclusions.tex =====
This study assessed turbulence models for simulating low-Prandtl-number
flows in channel, pipe, and backward-facing step configurations. The
models were evaluated in terms of reproducibility of reference
results, numerical robustness, and predictive accuracy against DNS and
high-fidelity data. A key outcome of this study is that several models
cannot be reliably assessed for low-Prandtl-number flows due to
limitations in reproducibility and numerical stability.

The KWSST-KAYS combination consistently provided accurate predictions of the
mean temperature field across all cases, although the underlying turbulent
heat flux is systematically underestimated (\Crefrange{subsec:channel}{subsec:bfs}). This accuracy
must therefore be attributed primarily to the very low Prandtl number of the
flows considered, rather than to an accurate representation of the turbulent
heat transport. In the BFS
configuration, a slight underestimation of the local Nusselt number was
observed downstream of reattachment, likely related to limitations of the underlying turbulence closure. Despite this, the $k$--$\omega$ SST model
accurately captured the mean velocity field in the recirculation region,
highlighting the robustness of this approach. The simplicity and local
nature of the Kays correlation further contribute to its computational
efficiency and robustness.

The AKN-MM combination demonstrated strong reproducibility, matching
reference results from the literature for channel, pipe, and BFS cases.
While minor discrepancies were observed in temperature variance, the
model provided consistent and accurate predictions of temperature and
Nusselt number distributions, making it a reliable option for
low-Prandtl-number flow simulations. 

The KLW-DAVIA combination reproduced trends similar to AKN-MM but
exhibited significantly reduced numerical robustness. Small variations
in the initial conditions of $\Omega$ led to divergence in channel
flow, contradicting the improved robustness claimed by
\citet{Manservisi2016}. In the BFS configuration, the thermal model
failed to converge entirely. This instability is likely associated with
nonlinear source terms involving $e^\Omega$ and $e^{\Omega_\theta}$.
As discussed in \Cref{subsec:channel}, this follows from the perturbation being
applied to the logarithmic variable $\Omega$ itself (rather than to $\omega$),
which corresponds to raising the physical dissipation rate to the tenth power
and strongly amplifies the exponential source terms $e^{\Omega}$ and
$e^{\Omega_\theta}$.
Given the similar predictive behaviour but inferior robustness, the
KLW-DAVIA combination does not provide advantages over AKN-MM.

The ShamsKE-AHFM combination showed inconsistent performance. While
the ShamsKE momentum model reproduced reference results, the coupled
thermal model failed to match published data and exhibited significant
deviations in Nusselt number predictions. These discrepancies are likely due to differences in model
implementation across CFD platforms and to the lack of a uniquely
defined formulation in the literature.

The EBRSM-AHFM combination reproduced reference results in channel flow
but showed limited numerical robustness: converged solutions were highly
sensitive to the absolute value of the kinematic viscosity and to the
initial Reynolds stress conditions, a behaviour attributed to the
numerical treatment rather than to the model formulation (\Cref{subsec:channel}).  In the BFS configuration, no
converged solution could be obtained, indicating limited applicability
of this approach. One possible source of the discrepancies observed for the EBRSM-AHFM
combination is the use of different EBRSM implementations. In the
present work, the OpenFOAM implementation based on \citet{Manceau2015}
is adopted, whereas \citet{Shams2019Number3} refer to a version based
on \citet{Manceau2014}. A future comparison using the same EBRSM
implementation, which is now available for OpenFOAM from
\citet{marocco_2024}, would help isolate the origin of these differences in channel and pipe
flow, and verify whether the same implementation resolves the convergence
issues observed in the backward-facing-step configuration.

A fundamental limitation of the AHFM formulation is its dependence on
the Reynolds-number-based coefficient $C_{t1}$, which introduces an inherent non-local dependency. The definition of the Reynolds number relies on
global reference quantities, which are not uniquely defined in complex
geometries, thereby limiting the general applicability of the model.

Overall, only a limited subset of models can be considered reliable for
low-Prandtl-number flows, as many approaches are affected by either
insufficient robustness or lack of reproducibility. The
KWSST-KAYS and AKN-MM combinations emerge as the most consistent and
reliable choices within the present assessment.

A summary of model performance in terms of reproducibility,
robustness, and recommended usage is provided in
\Cref{tab:overviewResults}. The present study highlights the importance of consistent model
formulation and implementation when assessing turbulence closures for
low-Prandtl-number heat transfer. The present findings apply to the
investigated Prandtl-number range $0.0088\le Pr\le 0.025$. No claim is made
regarding model performance at higher Prandtl numbers.

\begin{table}[!ht]
    \caption[Summary of model performance in low-Prandtl-number simulations]{Overview of model reproducibility, robustness, and recommendation for low-Prandtl-number flow simulations based on results from channel, pipe, and BFS cases.}\label{tab:overviewResults}
    \centering
    \begin{threeparttable}
        \begin{NiceTabular}{|p{3.8cm}|p{2.8cm}|p{2.2cm}|p{3.2cm}|}
            \hline
            \textbf{Model} & \textbf{Reproducibility} & \textbf{Robustness} & \textbf{Recommendation} \\
            \hline\hline
            KWSST-KAYS & Good & Good & Recommended \\
            AKN-MM & Good & Good & Recommended \\
            KLW-DAVIA & Good & Limited & Not recommended \\
            ShamsKE-AHFM & Limited & Good & Not recommended \\
            EBRSM-AHFM & Good & Limited & Further investigation \\
            \hline
        \end{NiceTabular}
    \end{threeparttable}
    
\end{table}

\FloatBarrier
% ===== end conclusions.tex =====

\section*{Acknowledgements}
% ===== begin acknowledgements.tex =====
Support by the German Research Foundation (DFG) under research project 423710075 is greatly acknowledged.  
% ===== end acknowledgements.tex =====

\section*{Author contributions}
L. Marocco: conceptualisation, methodology, supervision, writing -- original draft; J. Schmitt: software, validation, investigation; J. Neuhauser: supervision,  writing -- review \& editing; B. Frohnapfel: supervision, funding acquisition, project administration, writing -- review \& editing; D. Gatti: supervision, writing -- original draft. 

\section*{Declaration of generative AI and AI-assisted technologies in the manuscript preparation process}

During the preparation of this work, the author L. Marocco used ChatGPT (OpenAI) for spell-checking and for grammar and style improvements. After using this tool, the author reviewed and edited the content as needed. All authors reviewed the final manuscript and take full responsibility for the content of the published article.

%\appendix
% \appendixpage
%\section{Appendix}
\appendix
\section{Implementation of thermal turbulence models in \mbox{OpenFOAM}}
\label{sec:appendix_implementation}
% ===== begin appendix.tex =====
% \subsection{Implementation of thermal turbulence models in OpenFOAM}

The thermal turbulence models have been implemented within 
\ofversion{} using the abstract base class
\texttt{ThRASModel}, which provides a unified interface for all
thermal closures. The use of the template parameter
\texttt{BasicTurbulenceModel} ensures compatibility with both
incompressible and compressible momentum turbulence models.

All thermal models are derived from this base class and are selected
at run-time through the standard OpenFOAM run-time selection
mechanism. This design enables a modular coupling between momentum and
thermal turbulence models, allowing different combinations to be
systematically tested within a consistent numerical framework.

In addition to the models considered in the present study, further
implementations such as \texttt{laminar}, \texttt{AHFM2005}, and
\texttt{AHFMcc} are available but are not investigated here. The
\texttt{laminar} model corresponds to $\alpha_t = 0$.
The implemented models are integrated into the steady-state solvers
\texttt{simpleFoamThRAS} and
\texttt{chtMultiRegionSimpleFoamThRASSolid}. The former extends the
standard \texttt{simpleFoam} solver by including the energy equation
and thermal turbulence closures, while the latter enables simulations
with separate fluid and solid regions, allowing for the treatment of
conjugate heat transfer problems.

\begin{figure}[!ht]
\definecolor{blueDoxygen}{RGB}{25,25,112}
\definecolor{GrayDoxygen}{RGB}{191,191,191}

\footnotesize
\centering

\begin{tikzpicture}[
    scale=0.75,
    every node/.style={align=center,rectangle,draw=black},
    arrow/.style={-{Triangle[angle=45:3pt]},blueDoxygen,thick}
]

% Nodes
\node at (-14,0) (root) [draw=red]{IOdictionary};
\node at (-9.7,0) (ThRASModel) [fill=GrayDoxygen]
{ThRASModel \\ \textless BasicTurbulenceModel\textgreater};

\node at (-3.5,5) (laminar) {laminar \\ \textless BasicTurbulenceModel\textgreater};
\node at (-3.5,3) (RA) {ReynoldsAnalogy \\ \textless BasicTurbulenceModel\textgreater};
\node at (-3.5,1) (Kays) {Kays \\ \textless BasicTurbulenceModel\textgreater};
\node at (-3.5,-1) (MM) {MM \\ \textless BasicTurbulenceModel\textgreater};
\node at (-3.5,-3) (Via) {Via \\ \textless BasicTurbulenceModel\textgreater};
\node at (-3.5,-5) (AHFM2005) {AHFM2005 \\ \textless BasicTurbulenceModel\textgreater};

\node at (2.7,-4) (AHFMcc) {AHFMcc \\ \textless BasicTurbulenceModel\textgreater};
\node at (2.7,-6) (AHFMNRG) {AHFMNRG \\ \textless BasicTurbulenceModel\textgreater};

% Edges
\draw[arrow] (root) -- (ThRASModel);
\draw[arrow] (ThRASModel) to[out=70,in=-170] (laminar);
\draw[arrow] (ThRASModel) to[out=50,in=-175] (RA);
\draw[arrow] (ThRASModel) -- (Kays);
\draw[arrow] (ThRASModel) -- (MM);
\draw[arrow] (ThRASModel) to[out=-50,in=175] (Via);
\draw[arrow] (ThRASModel) to[out=-70,in=170] (AHFM2005);
\draw[arrow] (AHFM2005) -- (AHFMcc);
\draw[arrow] (AHFM2005) -- (AHFMNRG);

\end{tikzpicture}

\caption{Class hierarchy of the \texttt{ThRASModel} framework for
thermal turbulence models in OpenFOAM. The names appearing in the
diagram are the implementation names used in the code and may differ
from the model abbreviations adopted in the paper, such as
\texttt{Via} for DAVIA and \texttt{AHFMNRG} for AHFM.}
\label{fig:ThRASModel}
\end{figure}
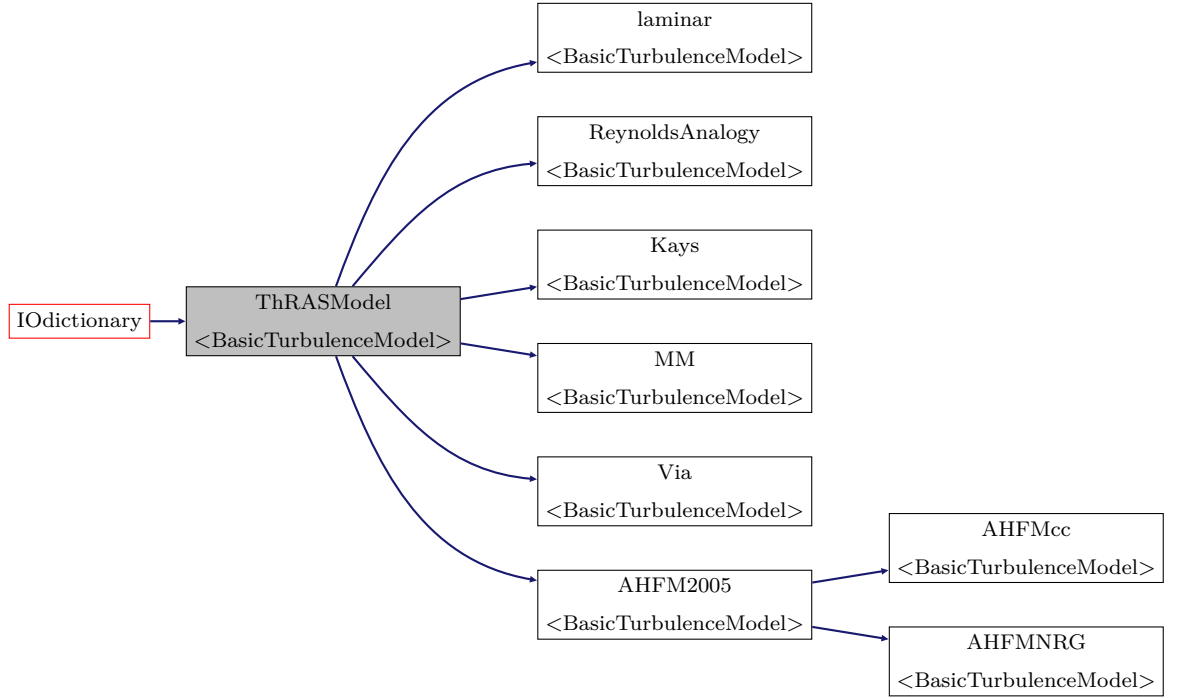

A schematic representation of the class hierarchy is shown in
\Cref{fig:ThRASModel}.

% ===== end appendix.tex =====

\FloatBarrier
% \section*{References}
% \label{References}

%% References

% \bibliographystyle{vancouver}
\bibliographystyle{model1-num-names}

\end{document}